\documentclass[11pt]{article}
\usepackage[letterpaper,textwidth=168mm,centering,vmargin=2.2cm]{geometry}
\usepackage[T1]{fontenc}
\usepackage{lmodern}
\usepackage{amsmath}
\usepackage{amssymb}
\usepackage{graphicx}
\usepackage{tgheros}
\usepackage{tikz}
\usepackage{soul}
\definecolor{cREF}{HTML}{000000}\definecolor{cREFfill}{HTML}{EDEDED}\definecolor{cEXACT}{HTML}{0072B2}\definecolor{cCUDA}{HTML}{E69F00}\definecolor{cNEU}{HTML}{999999}\definecolor{cGREEN}{HTML}{009E73}
\usetikzlibrary{arrows.meta,positioning,calc}
\usepackage{pgfplots}
\pgfplotsset{compat=1.18}
\usepackage{array}
\usepackage{booktabs}
\usepackage{caption}
\usepackage[numbers,super,sort&compress]{natbib}
\usepackage[hidelinks]{hyperref}
\providecommand{\eprint}[2][]{Preprint at \url{https://arxiv.org/abs/#2}}
\newcolumntype{L}[1]{>{\raggedright\arraybackslash}p{#1}}
\title{An exact and differentiable rewrite of a coupled weather and fire model}
\author{John S. Schreck$^{1,*}$, John Dennis$^{1}$, Iris Linck$^{1}$, Maria Frediani$^{2}$,\\ Pedro A. Jim\'enez Mu\~noz$^{2}$, Matthew Rehme$^{1}$, Nihanth Cherukuru$^{1}$, and David John Gagne II$^{1}$\\[1ex]
\small $^{1}$Computational and Information Systems Laboratory,\\
\small $^{2}$Research Applications Laboratory,\\
\small NSF National Center for Atmospheric Research, Boulder, CO, USA\\
\small $^{*}$Corresponding author: schreck@ucar.edu}
\date{}

\begin{document}
\maketitle

\section*{Abstract}

\noindent AI is changing weather and climate prediction by replacing physical models with learned ones. Here we show a different path, in which AI coding agents rebuild the physical model itself in a machine-learning framework and reproduce its forecast bit for bit. We reconstruct selected configurations of the Weather Research and Forecasting model and its coupled fire physics in PyTorch, treating every difference from the corrected Fortran code as a bug. Across twelve hours of a Marshall Fire simulation, the two independently advancing codes agree bit for bit at all 365 checkpoints, in the atmosphere, fire and firebrand particles. This agreement holds when the code runs one operation at a time, with the Fortran code's mathematical library. On a separate fast route, the reconstructed model runs about five times faster on one GPU than the double-precision Fortran code on a 128-core CPU node. One backward pass maps the fire's sensitivity to the wind at every grid point. In synthetic tests, its derivatives show how to change wind and fuel moisture to correct a misplaced fire front. Neural networks replace the radiation and fire-spread calculations in 24-hour coupled forecasts, and we measure their effect against the physical calculation. An established model thus becomes fast, differentiable and open to learned components, with its tested calculation intact.

\par

\section*{Introduction}
Machine learning is changing weather and climate prediction \cite{bi2023pangu, lam2023graphcast, price2025gencast, bodnar2025aurora, schreck2025credit, chapman2025camulator, wattmeyer2025ace2}. In medium-range forecasting, models learned from data now match or exceed the skill of the leading physical forecast systems, and they run in minutes on graphics processing units (GPUs) \cite{bi2023pangu, lam2023graphcast, price2025gencast, bodnar2025aurora, schreck2025credit}. These models are trained on reanalyses produced with physical models, however, and they set aside the equations and physical schemes in which decades of scientific knowledge are held \cite{EasterbrookJohns2009}. The physical models keep that knowledge, but most are large Fortran codes that were not written for this hardware. Bringing established models onto GPUs would put that computing power to work on forecasts built from decades of tested physics.

Much of what scientists now want from a physical model is hard to obtain from the Fortran code. Ensemble forecasts require many simulations in the time available for a decision, and GPUs can make each one faster \cite{Schulthessetal2019, Baueretal2021}. Variational data assimilation, which estimates poorly known winds or fuel moisture from observations, needs derivatives of the forecast with respect to its inputs \cite{ledimet1986variational}. A neural network proposed to replace a physical scheme can be judged only inside the coupled model \cite{wang2022neuralphysics, kochkov2024neuralgcm}. Machine-learning frameworks such as PyTorch \cite{paszke2019pytorch} provide GPU execution, automatic differentiation and a common interface for physical and learned components \cite{shen2023differentiable}. The test of a rewrite is whether it can deliver these capabilities and still reproduce the established calculation.

Moving an established model into such a framework restructures the software. Scalar loops become parallel array operations and state is reorganized \cite{Dahmetal2023}. These changes can alter the numerical calculation, and plausible weather is a weak test for a translation error whose effects amplify with time \cite{RosinskiWilliamson1997, baker2015cect}. The decisive test is to reproduce the established calculation exactly \cite{Arteagaetal2014, Donahueetal2024}. When we replace a radiation scheme with a neural network \cite{zhong2023wrfml}, for example, we can compare the resulting forecast with one that uses the physical scheme in the same model.

The Weather Research and Forecasting model (WRF) provides a demanding test \cite{skamarock2019wrf, powers2017wrf}. We use its coupled fire component, WRF-Fire \cite{mandel2011sfire, coen2013wrffire}, and simulate the December 2021 Marshall Fire, Colorado's costliest wildfire to date \cite{wwa2022marshall}. Fire heat changes the atmospheric flow, which in turn changes fire spread and the transport of firebrands, or embers, that can ignite new fires \cite{frediani2025spotting}. A difference introduced in one component therefore reaches the others later, so an exact copy must reproduce every component and every exchange between them.

Large language models (LLMs) have translated individual pieces of scientific software \cite{ranasinghe2025fortrancpp}, including into differentiable frameworks \cite{zhou2024chatgptjax, lahlou2026clmml}. LLM-Assisted Fortran Translation (LAFT) combined translation with numerical validation and performance profiling for a microphysics scheme \cite{linck2026laft}. Separate derivative codes for WRF (tangent-linear and adjoint models) support variational data assimilation, although with simplified physics \cite{zhang2013wrfplus, wrfda2017guide, wrfda2026overview}. Learned components have been coupled to physical models \cite{zhong2023wrfml, kochkov2024neuralgcm}. Coupled model ports to the JAX framework have reported statistical consistency \cite{davenport2026jcm, yuan2026wrfjax}. To our knowledge, no rewritten weather model has reproduced the trajectory of its Fortran code bit for bit.

Here we show that AI coding agents can rebuild an established physical model inside a machine-learning framework without changing its calculation. We develop and evaluate WRF-Speed, a PyTorch reconstruction of selected coupled WRF-Fire configurations (Methods and Supplementary Table~\ref{S-tab:evidence-scope}), written with AI coding agents and checked against compiled Fortran references. In its exact route, which runs one operation at a time, exactly as written, WRF-Speed agrees bit for bit with a corrected Fortran reference over twelve hours of a nested Marshall Fire simulation, with each code advancing its own forecast state. A separate fast route, which uses standard GPU arithmetic and gives up bit-for-bit agreement, runs the nested calculation on one GPU about five times faster than the double-precision Fortran code on a 128-core CPU node. One backward pass through the coupled Marshall calculation maps how the fire's heat release responds to the wind at about 11 million grid values. WRF-Speed's checked derivatives correct a misplaced fire front in idealized synthetic tests. Neural networks trained to reproduce the physical schemes replace the radiation and fire-spread calculations in 24-hour coupled forecasts, which we compare with the physical forecasts. The established model thus becomes fast, differentiable and open to learned components at once, with its tested calculation intact.

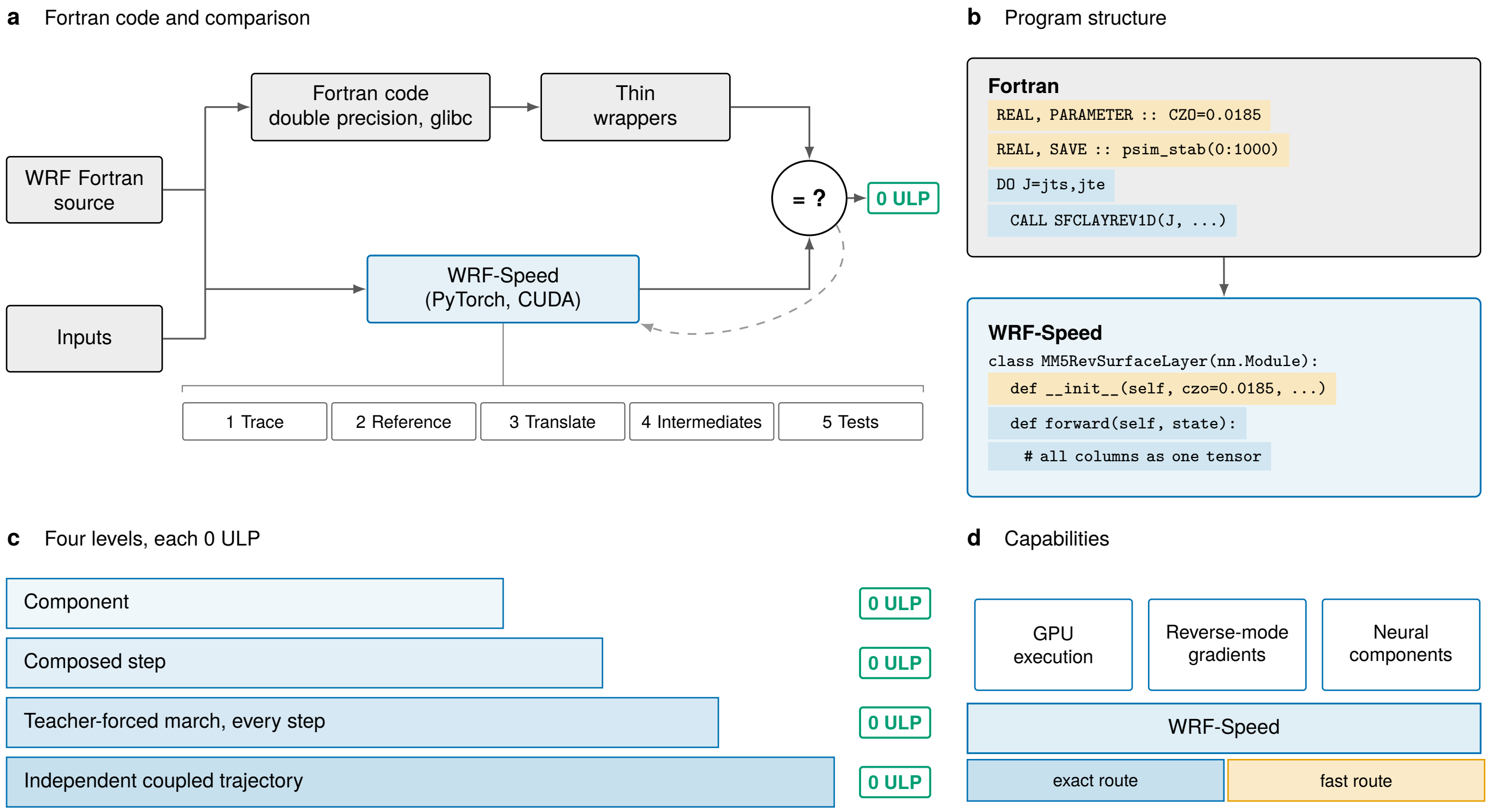
\begin{figure}[!t]
\centering
\makebox[\textwidth][c]{
\begin{tikzpicture}[x=1mm,y=1mm,
  font=\sffamily\fontsize{7}{8.4}\selectfont,
  ref/.style={draw=cREF, line width=0.5pt, fill=cREFfill, rounded corners=1.2pt, align=center, inner sep=1.4mm, minimum height=8mm, text width=#1},
  ex/.style={draw=cEXACT, line width=0.6pt, fill=cEXACT!10, rounded corners=1.2pt, align=center, inner sep=1.4mm, minimum height=8mm, text width=#1},
  chip/.style={draw=cREF!60, line width=0.4pt, fill=white, rounded corners=0.8pt, align=center, inner sep=0.6mm, minimum height=4.5mm, text width=16.2mm, font=\sffamily\fontsize{6}{7.2}\selectfont},
  cmp/.style={draw=cREF, line width=0.6pt, fill=white, circle, minimum size=9mm, inner sep=0},
  ok/.style={draw=cGREEN, line width=0.7pt, fill=white, rounded corners=1pt, inner sep=1mm, font=\sffamily\bfseries\fontsize{6.5}{7.8}\selectfont, text=cGREEN},
  tier/.style={draw=cEXACT, line width=0.5pt, fill=cEXACT!#1, minimum height=6mm, anchor=west, align=left, inner xsep=2mm},
  arr/.style={-{Latex[length=1.6mm,width=1.2mm]}, line width=0.6pt, draw=cREF!65},
  lab/.style={font=\sffamily\bfseries\fontsize{8}{9.6}\selectfont, anchor=base west, inner sep=0},
  ttl/.style={font=\sffamily\fontsize{7}{8.4}\selectfont, anchor=base west, inner sep=0}]
\node[lab] at (0,96) {a}; \node[ttl] at (4.5,96) {Fortran code and comparison};
\node[ref=16mm] (src) at (9.4,76) {WRF Fortran\\source};
\node[ref=16mm] (inp) at (9.4,58) {Inputs};
\node[ref=26mm] (orc) at (44,86) {Fortran code\\double precision, glibc};
\node[ref=20mm] (wrp) at (76,86) {Thin\\wrappers};
\node[ex=30mm] (prt) at (60,64) {WRF-Speed\\(PyTorch, CUDA)};
\node[cmp, font=\sffamily\bfseries\fontsize{8}{9.6}\selectfont] (eq) at (97,75) {= ?};
\node[ok, anchor=west] (ok) at (104,75) {0 ULP};
\draw[line width=0.6pt, draw=cREF!65] (src.east) -- (24,76); \draw[line width=0.6pt, draw=cREF!65] (inp.east) -- (24,58);
\draw[line width=0.6pt, draw=cREF!65] (24,58) -- (24,86);
\draw[arr] (24,86) -- (orc.west);
\draw[arr] (24,64) -- (prt.west);
\draw[arr] (orc) -- (wrp);
\draw[arr] (wrp.east) -| (eq.north);
\draw[arr] (prt.east) -| (eq.south);
\draw[arr] (eq) -- (ok);
\draw[arr, dashed, draw=cNEU] (eq.south east) to[out=-60,in=-20] (prt.south east);
\foreach \i/\t in {0/Trace,1/Reference,2/Translate,3/Intermediates,4/Tests} {\node[chip] (g\i) at (30+\i*18,48) {\number\numexpr\i+1\relax~\t};}
\draw[line width=0.4pt, draw=cREF!50] ([yshift=1.2mm]g0.north west) -- ++(0,0.8) -- ([yshift=2mm]g4.north east) -- ([yshift=1.2mm]g4.north east);
\draw[line width=0.4pt, draw=cREF!50] (prt.south) -- (prt.south |- 0,{48+2.25+2});
\node[lab] at (116,96) {b}; \node[ttl] at (120.5,96) {Program structure};
\node[draw=cREF, line width=0.5pt, fill=cREFfill, rounded corners=2pt, anchor=north west, minimum width=62mm, minimum height=24mm, align=left, inner sep=1.6mm, text width=57mm] (ftn) at (116,92)
 {\textbf{Fortran}\\[0.6mm]{\ttfamily\fontsize{6}{7.4}\selectfont
 \colorbox{cCUDA!25}{REAL, PARAMETER :: CZO=0.0185}\\
 \colorbox{cCUDA!25}{REAL, SAVE :: psim\_stab(0:1000)}\\
 \colorbox{cEXACT!18}{DO J=jts,jte}\\
 \colorbox{cEXACT!18}{\ \ CALL SFCLAYREV1D(J, ...)}}};
\node[draw=cEXACT, line width=0.6pt, fill=cEXACT!8, rounded corners=2pt, anchor=north west, minimum width=62mm, minimum height=24mm, align=left, inner sep=1.6mm, text width=57mm] (pyt) at (116,63)
 {\textbf{WRF-Speed}\\[0.6mm]{\ttfamily\fontsize{6}{7.4}\selectfont
 class MM5RevSurfaceLayer(nn.Module):\\
 \colorbox{cCUDA!25}{\ \ def \_\_init\_\_(self, czo=0.0185, ...)}\\
 \colorbox{cEXACT!18}{\ \ def forward(self, state):}\\
 \colorbox{cEXACT!18}{\ \ \ \ \# all columns as one tensor}}};
\draw[arr] (ftn.south) -- (pyt.north);
\node[lab] at (0,33) {c}; \node[ttl] at (4.5,33) {Four levels, each 0 ULP};
\foreach \k/\t/\w/\f in {0/Component/60/6,1/Composed step/72/11,2/{Teacher-forced march, every step}/86/16,3/Independent coupled trajectory/100/22} {
  \path[draw=cEXACT, line width=0.5pt, fill=cEXACT!\f] (0,{23-\k*7.2}) rectangle (\w,{29-\k*7.2});
  \node[anchor=west, align=left, inner sep=0] at (2,{26-\k*7.2}) {\t};
  \node[ok, anchor=west] at (103,{26-\k*7.2}) {0 ULP};
}
\node[lab] at (116,33) {d}; \node[ttl] at (120.5,33) {Capabilities};
\foreach \x/\t in {126.5/{GPU\\execution},147.5/{Reverse-mode\\gradients},168.5/{Neural\\components}} {
  \node[draw=cEXACT, line width=0.5pt, fill=white, rounded corners=1pt, minimum width=19mm, minimum height=11mm, align=center, font=\sffamily\fontsize{6.5}{7.8}\selectfont] at (\x,21) {\t};
}
\node[draw=cEXACT, line width=0.6pt, fill=cEXACT!12, minimum width=62mm, minimum height=6mm, anchor=north west] at (116,14) {WRF-Speed};
\node[draw=cEXACT, line width=0.5pt, fill=cEXACT!22, minimum width=31mm, minimum height=5mm, anchor=north west, font=\sffamily\fontsize{6}{7.2}\selectfont] at (116,7.2) {exact route};
\node[draw=cCUDA, line width=0.5pt, fill=cCUDA!22, minimum width=31mm, minimum height=5mm, anchor=north west, font=\sffamily\fontsize{6}{7.2}\selectfont] at (147.5,7.2) {fast route};
\end{tikzpicture}

}
\caption{\textbf{WRF-Speed is rewritten from the WRF Fortran code and checked at 0~ULP.} ULP denotes a unit in the last place, the smallest possible difference between two floating-point numbers. Agreement at 0~ULP means the two values are identical in every bit. \textbf{a}, Translation from the Fortran code. When a result differs, we search operation by operation for the first one that disagrees (dashed loop). Numbered boxes mark the five stages used for each scheme. \textbf{b}, Corresponding surface-layer code, with matching operations highlighted. \textbf{c}, Four levels of comparison. A teacher-forced march starts every step from the Fortran state. \textbf{d}, Capabilities and arithmetic routes: the exact route (blue) for comparison with the Fortran code and the fast route (orange) for timing, differentiation and the neural-network forecasts.}
\label{fig:concept}
\end{figure}

\section*{Results}
\subsection*{Bitwise agreement in component tests}
Tests of individual model components (e.g., physics schemes) show that WRF-Speed reproduces the compared Fortran outputs bit for bit using its exact arithmetic route. The PyTorch implementation reorganizes the Fortran nested loops over independent grid points into array operations that GPUs can run in parallel. Calculations that depend on one another retain their order (Fig.~\ref{fig:concept}b).

WRF-Speed runs in two ways. The exact arithmetic route executes operations without compilation and uses the same mathematical library as the Fortran reference. A separate fast route uses standard GPU arithmetic and combines the most expensive operations into compiled GPU kernels, allowing differences in the last bits of some results. Both routes support automatic differentiation to calculate the sensitivity of model outputs to selected inputs. We evaluate the accuracy and applicability of these derivatives separately.

The component comparisons cover 29 output fields from the surface-layer scheme, 42 from the Noah land-surface model and 12 from the YSU boundary-layer scheme. They also cover the 14 outputs of the WSM6 microphysics routine, the 13 flux and heating-rate outputs of the RRTMG longwave radiation component and the level-set function of the fire calculation over 36 consecutive timesteps. A test with 200 particles covers all eight outputs of the firebrand burnout, terminal velocity, removal and deposition calculations. Optimized and unoptimized builds of the Fortran firebrand code also produce bitwise identical outputs in these tests.

In six idealized WRF test cases, WRF-Speed also reproduces every compared Fortran field over integrations of 3 to 12 timesteps. These short integrations extend the comparison from individual components to their combined execution. Supplementary Table~\ref{S-tab:inventory} lists the components used in the reported Marshall calculation.

\subsection*{Bitwise agreement in the coupled Marshall simulation}
\begin{figure}[p]
\centering
\makebox[\textwidth][c]{\includegraphics{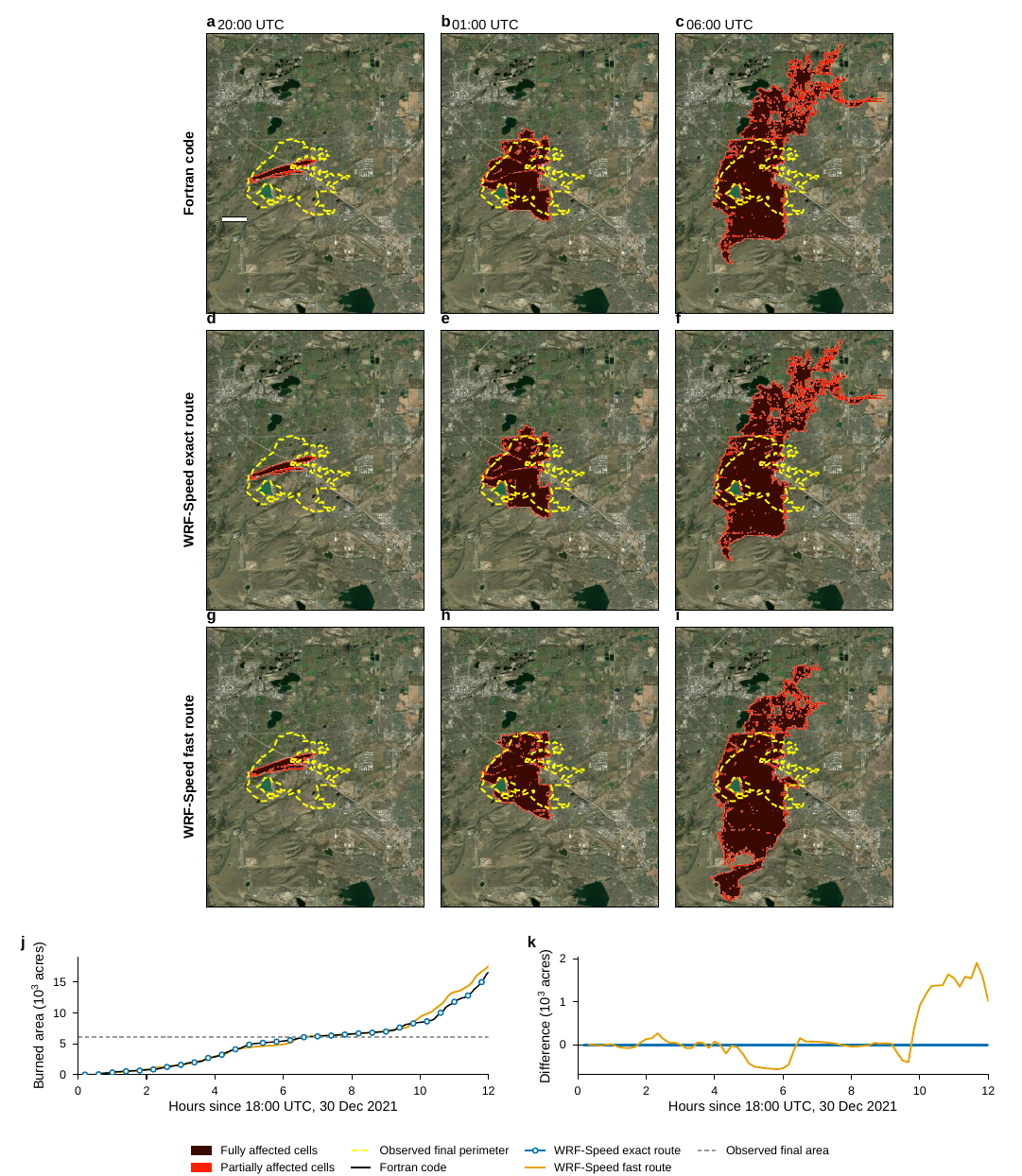}}
\caption{\textbf{The WRF-Speed exact route reproduces the Marshall Fire trajectory of the Fortran code bit for bit.} \textbf{a--i}, Fully affected (dark red, burned fraction at least 0.999) and partially affected (bright red) fire cells for the Fortran code (a--c), the WRF-Speed exact route (d--f) and the separate fast-route run (g--i). Columns show 20:00 UTC on 30 December and 01:00 and 06:00 UTC on 31 December 2021. Yellow dashes mark the observed final perimeter. Scale bar, 2\,km. \textbf{j}, Burned area for the Fortran code (black), the exact route (blue circles) and the fast-route run (orange). Dashed line, observed final area. \textbf{k}, Burned area of each WRF-Speed run minus that of the Fortran code. The exact-route difference (blue) is zero at every checkpoint. Basemap, Esri World Imagery.}
\label{fig:parity}
\end{figure}

WRF-Speed reproduces the atmospheric and fire fields of the Marshall Fire simulation over twelve simulated hours (Fig.~\ref{fig:parity}), using the corrected double-precision Fortran code and the exact arithmetic route. We compare winds, temperature, pressure and moisture on the 1-km parent and 111-m nested domains, together with the state of the fire at 2-minute intervals. Across 365 checkpoints, all 22,686,546,895 floating-point values compared by their bit patterns match. Arrays with different storage types also agree numerically, bringing the total to about 68 million array values per checkpoint. The physical properties of active firebrands agree at every comparison. Each code advances its own atmospheric and fire state, so the comparison validates the coupling and feedback as the simulation evolves.

This agreement shows that the rewrite preserves the tested coupled calculation, including the interactions that component tests can miss. Both codes overestimate the fire area when compared with its observed final perimeter. The separate fast-route run uses standard GPU arithmetic and follows a different trajectory (Fig.~\ref{fig:parity}g--i,k). We examine sensitivity to small numerical changes below by perturbing the winds within a single implementation.

The spotting scheme releases particles from burning cells, transports them following the three-dimensional wind, calculates combustion and deposition, and ignites new fires when landing conditions are met. Spotting accelerates the spread of a fire and allows it to jump firebreaks, and it plays a dominant role in wind-driven fires. The Fortran implementation exchanges particle arrays among MPI ranks (the separate processes that share the work on a CPU cluster) and must preserve particle state across restarts. The physical process and its numerical model representation are highly nonlinear. A single last-bit difference in a particle's release height or fall velocity can change the location of a new ignition, leading to a substantial divergence in the state of two runs. Thus, reproducing these events requires consistent particle calculations and state transfers.

The firebrand count reaches 87,093 active particles, and their corresponding physical properties agree with the Fortran code at every comparison (Supplementary Fig.~\ref{S-fig:edseries}b).
More than half of the corrections to the Fortran code concern the firebrand calculations (Supplementary Table~\ref{S-tab:corrections}). For example, an uninitialized variable on ranks without particles caused the Fortran results to differ between runs.

Comparisons of the coupled model found errors in the transfer of values that the individual component tests missed. In one case, the nested domain recomputed its own ozone profile instead of using the one interpolated from its parent. The downward longwave (infrared) radiation changed by at most $2.7\times10^{-5}$ in relative terms, yet after 34 parent steps more than half of the atmospheric values (136 of 246 million) differed. In another case, the radiation component missed a column-mass factor when passing the heating rates to the dynamical core, resulting in a radiative heating about $10^{5}$ times too small, although the radiation routine matched the Fortran code. The coupled model must therefore be tested as a whole, because an error in the connecting code can change the entire forecast while every component still matches.

\subsection*{GPU performance and derivative checks}
Writing the calculation as array operations lets the GPU process independent grid cells and atmospheric columns in parallel. A fast forecast must also avoid temporary arrays and the cost of starting many small GPU operations. We therefore combine selected calculations into larger GPU kernels (small programs that run on the GPU), written in CUDA and Triton\cite{tillet2019triton} or generated by the PyTorch compiler\cite{ansel2024pytorch2}. We record each step's sequence of kernels once and replay it as a unit, called a CUDA graph. A replay repeats the same sequence of operations on the current model state, so each replayed step still calculates a new state. Timesteps that update radiation run without replay.

\begin{figure}[!t]
\centering
\makebox[\textwidth][c]{\includegraphics{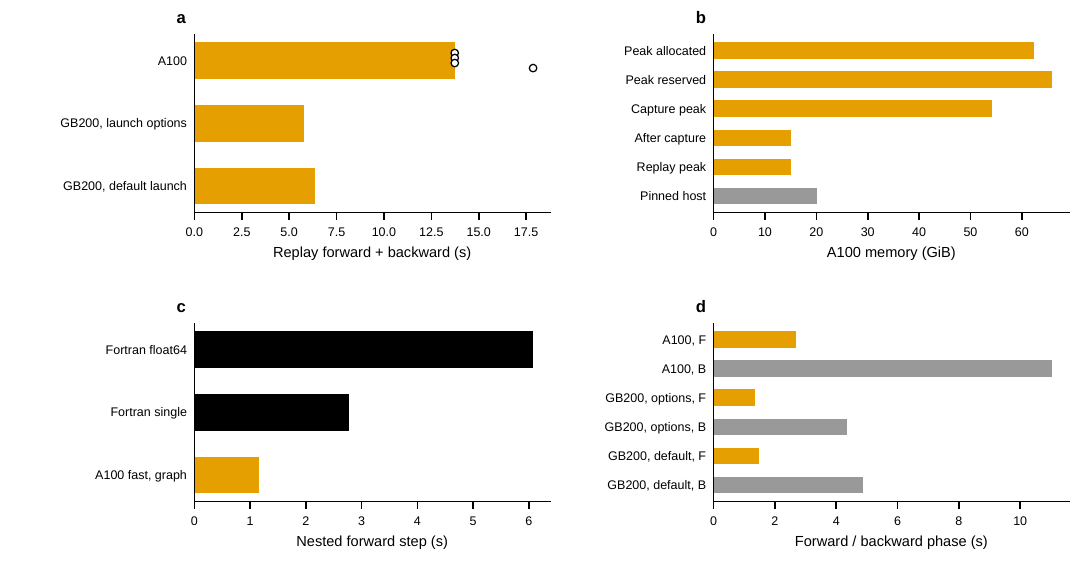}}
\caption{\textbf{On one GPU, the fast route runs a nested forecast step about five times faster than the double-precision Fortran code.} \textbf{a}, Median wall time for a forward and backward pass through nine substeps (3 simulated seconds) of the 111-m nest, with parent forcing and radiation tendencies held fixed. Medians cover four A100 or eight GB200 repeats, and open circles show the four A100 repeats. The two GB200 bars are runs with and without a set of launch options. \textbf{b}, A100 memory during that pass. Orange bars show GPU memory and the gray bar shows the intermediate arrays held in CPU memory. \textbf{c}, Mean wall time per nested parent step, forward only and including radiation updates, for the double-precision and single-precision Fortran builds on a 128-core CPU node (black) and for WRF-Speed on one A100 (orange). \textbf{d}, Forward (F, orange) and backward (B, gray) parts of the pass in a. The GB200 forward and backward times come from a separate run from the wall times in a. All GPU timings use the fast route.}
\label{fig:capability}
\end{figure}

Over 100 parent steps of the nested forecast, including five radiation updates, the fast route averages 1.166\,s per step on one A100 80\,GB GPU in double precision, as the mean of two runs (Fig.~\ref{fig:capability}c). Steps without a radiation update average 1.118\,s, and steps with one average 2.084\,s. On a 128-core CPU node, Fortran averages 6.075\,s in double precision and 2.769\,s in the single precision used in production. Over this five-minute simulated period, GPU compute throughput is 5.2 times that of double-precision Fortran and 2.4 times that of production single-precision Fortran. The two codes start from separately produced restart states, and the timings exclude startup and output. The final WRF-Speed state matches the saved state from an earlier run without replay, byte for byte.

We also differentiate nine fire-coupled substeps on the 111-m nest, holding the forcing from the parent and the radiation heating rates fixed. We measure how the mean moist potential temperature at the end of those three seconds changes with uniform offsets to the initial wind and temperature. A forward pass advances the model and a backward pass calculates the derivatives. Together the passes take 13.73\,s on an A100, where we store selected intermediate states in CPU memory (Fig.~\ref{fig:capability}a,b). With a different software build and intermediate states kept on the GPU, a GB200 run takes 5.76\,s. In both configurations the state, output and derivatives agree bit for bit with the corresponding uncompiled PyTorch calculation. This agreement checks the fast implementation, not the accuracy of the derivatives.

We test the accuracy of the derivatives in a separate experiment, by changing an input slightly, rerunning the model and comparing the change in the output with the derivative (a finite-difference check). In one Marshall nest substep, the derivative of burned area with respect to wind misses the tolerance at all five wind increments chosen in advance. In follow-up runs it meets the tolerance at each of five smaller increments, from 0.00001 to 0.005\,m\,s$^{-1}$. It also meets the tolerance at four of the original increments when we hold the model's conditional choices fixed, i.e., the places where the code selects between alternatives, such as which side of a grid cell the wind arrives from (Methods).

\subsection*{Fire sensitivity to the three-dimensional wind}
We next differentiate the coupled Marshall simulation with respect to the wind at every grid point. We start the 111-m nest at 19:15 UTC on 30 December 2021, shortly after a spot fire crossed US Highway 36, and advance nine substeps (3 simulated seconds) with the atmosphere, fire, firebrands and fire--atmosphere feedback active. The outputs are the total heat release of the fire (0.89\,GW) and its burned area (1.60\,km$^2$) at the end of the 3\,s. The sensitivity at a grid point is the change in an output per 1\,m\,s$^{-1}$ change in the initial east--west or north--south wind at that point. Estimating these values by rerunning the model would require one simulation for each of the roughly 5.4 million values of each wind component. By contrast, reverse-mode differentiation (i.e., backpropagation in machine learning, or the adjoint method in data assimilation) gives all of them from one backward pass per output.

\begin{figure}[!t]
\centering
\includegraphics[width=\textwidth]{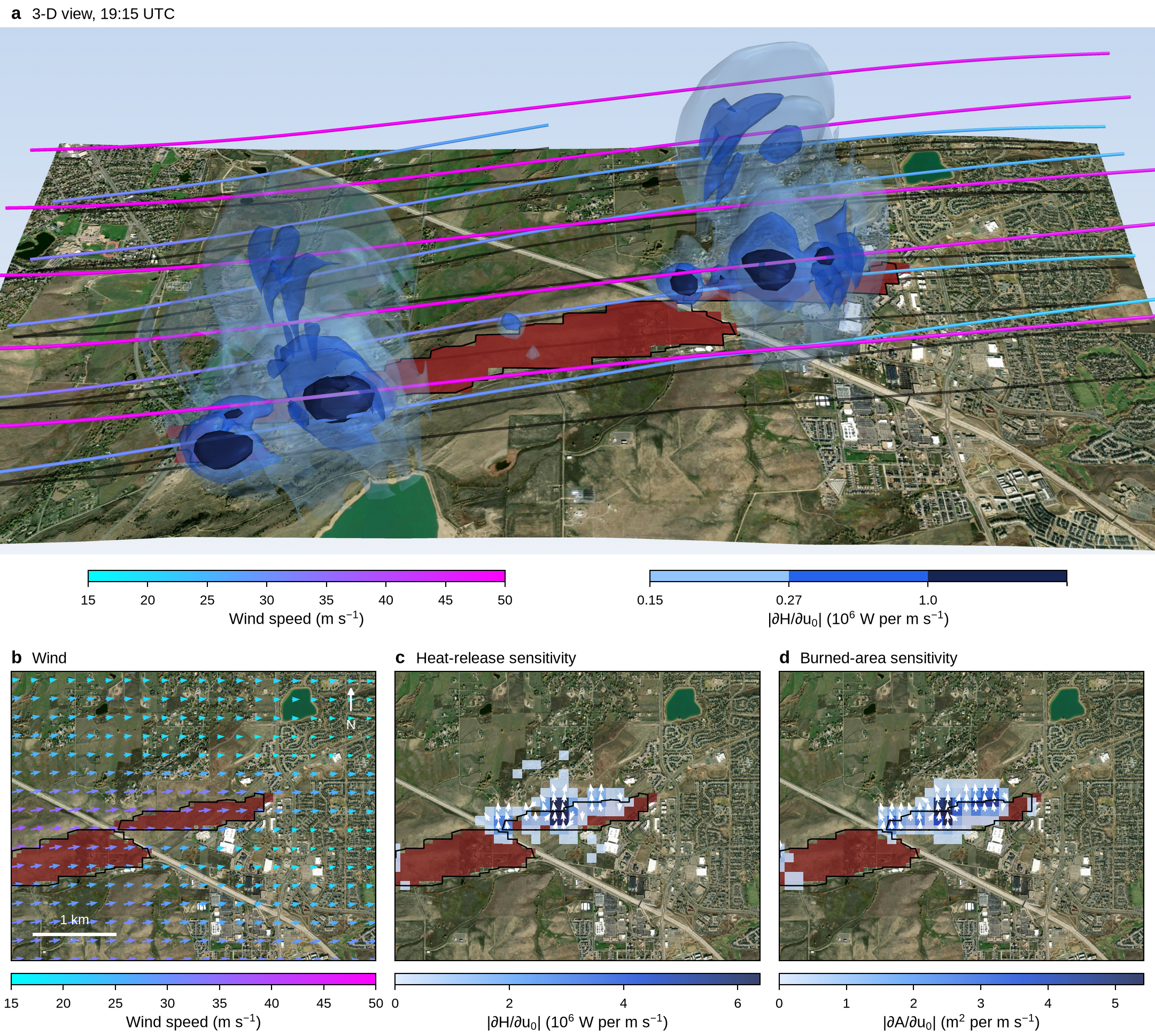}
\caption{\textbf{The heat release and burned area of the Marshall Fire respond to the wind throughout the three-dimensional grid.} Fields are from the 111-m nest at 19:15 UTC on 30 December 2021. Each sensitivity is the change in total heat release (0.89\,GW) or burned area (1.60\,km$^2$) 3 simulated seconds later per 1\,m\,s$^{-1}$ change in the initial horizontal wind at one grid point. \textbf{a}, Three-dimensional view with north up and 2.5-times vertical exaggeration. Blue shells enclose the largest 3\%, 0.7\% and 0.1\% of the heat-release sensitivity magnitudes. Streamlines follow the wind and are colored by horizontal speed (15--50\,m\,s$^{-1}$). The burned area is dark red with a black outline. \textbf{b}, Wind on the third model level. Arrows point downwind and are colored by speed on the same scale as in a. \textbf{c}, Magnitude of the heat-release sensitivity on the third model level (W per m\,s$^{-1}$). White arrows point in the direction in which added wind at that point would most increase heat release. \textbf{d}, As in c, for burned area (m$^2$ per m\,s$^{-1}$). Panels b--d are zoomed on the spot fire that crossed US Highway 36. Basemap, Esri World Imagery.}
\label{fig:windsens}
\end{figure}

Figure~\ref{fig:windsens} shows that the sensitivity is largest near the ground on the spot fire. For both outputs and both wind components, the maximum lies on the third model level at neighboring grid points (Fig.~\ref{fig:windsens}a,c,d). The north--south wind dominates there. The largest heat-release sensitivity is $2.06\times10^{7}$\,W per m\,s$^{-1}$ for the north--south wind (2.3\% of the total heat release per m\,s$^{-1}$ at a single grid point), compared with $4.18\times10^{6}$\,W per m\,s$^{-1}$ for the east--west wind. The corresponding burned-area values are 27.0 and 4.37\,m$^2$ per m\,s$^{-1}$. Around the spot fire, the white arrows point mostly north (Fig.~\ref{fig:windsens}c,d). In particular, the spot fire is a narrow strip aligned with the westerly flow (Fig.~\ref{fig:windsens}b), and a northward component turns the wind across its long northern flank toward unburned fuel.

Summed over the nest, the sensitivities give the response to a uniform change in the wind. Heat release increases by 31\,MW per m\,s$^{-1}$ of added eastward wind, which strengthens the existing westerly flow. The north--south sum is negative ($-42$\,MW per m\,s$^{-1}$), because the sensitivity changes sign across the nest and regions of the opposite sign outweigh the spot fire. The burned-area sums have the same signs ($+32.9$ and $-19.4$\,m$^2$ per m\,s$^{-1}$).

Note that about 38\% of the wind values affect the outputs within 3\,s, far beyond the burned area. A wind change travels through the model as pressure adjustments carried by sound waves and through the transport calculations, which reach several neighboring cells at each of the 27 Runge--Kutta stages. These distant values are small, and the shells that enclose the largest 3\% of the magnitudes stay above the fire and rise with its plumes (Fig.~\ref{fig:windsens}a). We have not compared the individual values with finite differences. The burned-area derivative on this nest agrees with finite differences only for wind changes of 0.005\,m\,s$^{-1}$ and below, or with the conditional choices of the code held fixed (Methods). Each value is thus a local rate over 3\,s rather than a prediction for a 1\,m\,s$^{-1}$ change. Even so, a single backward pass identifies where a wind observation or correction would most change the forecast fire.

\subsection*{Fire-front sensitivity to fuel moisture}
A derivative predicts how far a fire front moves when the fuel becomes drier. From a single forecast, it gives the change in front position for a small change in fuel moisture, which fresh simulations can then check. We use a simpler setting than the Marshall case so that the effect of a single input is easy to see without terrain, firebrands and full physics. The domain is flat and 2\,km across, and the model advances in 0.25-s steps. The fire and atmosphere remain coupled, and only the initial fuel moisture changes, with the dry fuel load held fixed. We track the head of the fire, its leading edge, along five fixed lines (Methods).

Starting from a forecast at 8\% moisture, the derivative predicts that the head moves faster at 7\% and slower at 9\%. After 60\,s (240 steps) the two changes shift the head by about a meter in opposite directions, and the prediction matches fresh simulations to within 0.09\,m (Fig.~\ref{fig:moisture}a--c). Central finite differences, which estimate the derivative from simulations with slightly higher and lower moisture, confirm the derivative itself at each of the three forecast times. Thus, one differentiated forecast gives both the drying and moistening responses.

The derivative's straight-line prediction holds only near the starting point. At 6\% and 10\% moisture its error grows to 16--18\% of the simulated change in head position (Supplementary Table~\ref{S-tab:moisture-response}), which shows that the response to moisture is curved. The experiment thus measures both the sensitivity to moisture and, for this case, how far from 8\% a single derivative remains useful.

\subsection*{Gradient-based correction of synthetic fire fronts}
Derivatives can also show how to change a forecast whose fire front is in the wrong place. Estimating winds or fuel moisture from observed fronts is established in wildfire modeling\cite{mandel2009wildfireda,mandel2012perimeter,rochoux2014parameter,rochoux2015state}. Here we calculate how each front position responds to changes in two initial wind offsets and fuel moisture. These derivatives form a matrix, the front Jacobian, with one row for each front position and one column for each input. We begin from two different initial guesses of the wind offsets and moisture, which we call starting points. We fit the three inputs to noisy synthetic fronts at 120 and 180\,s, generated with a different version of Fortran WRF running at half our timestep. Before the update, we check the derivatives of front position and of the misfit against finite differences.

\begin{figure}[!t]
\centering
\includegraphics[width=\textwidth]{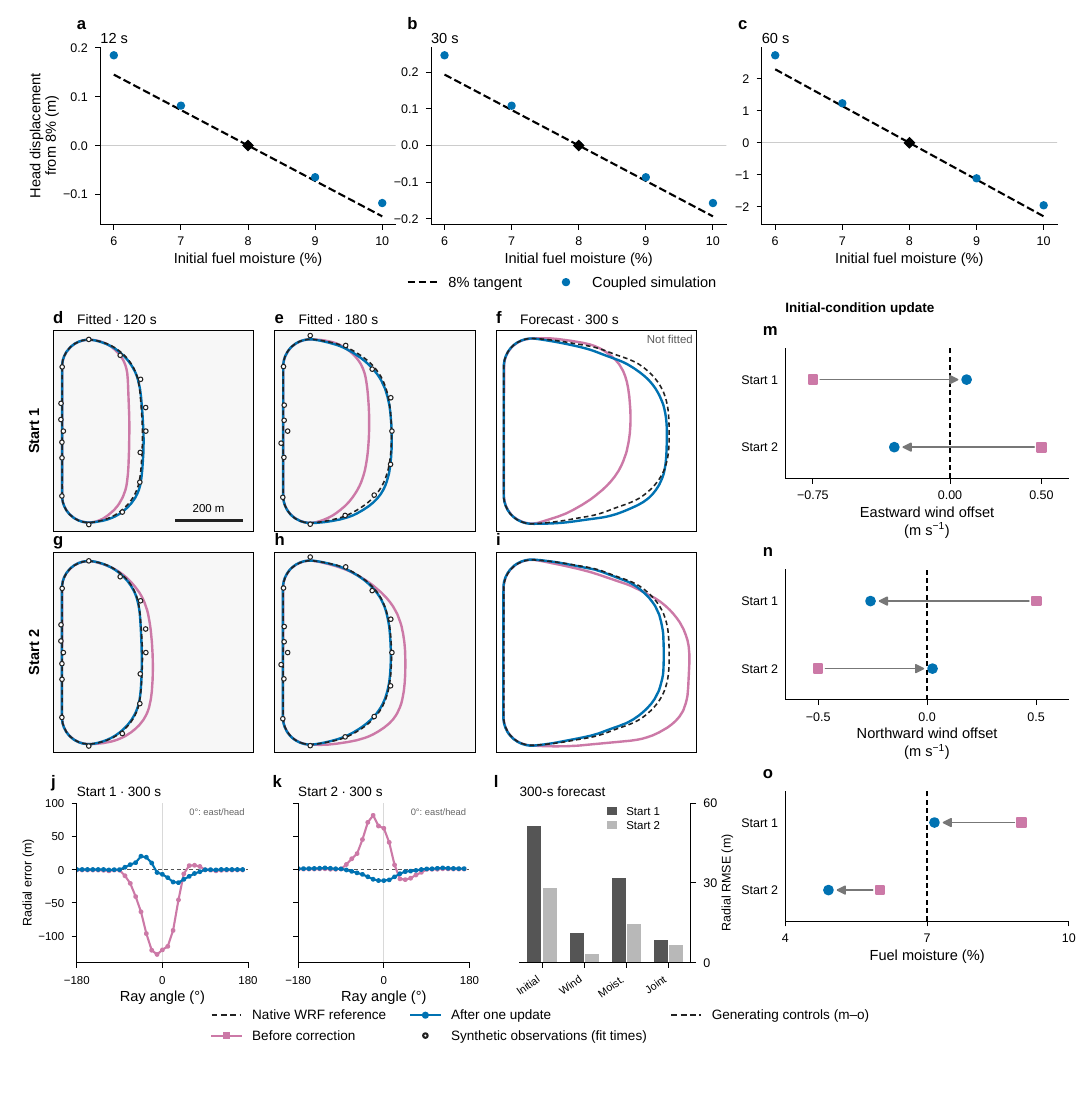}
\caption{\textbf{Derivatives predict the fire response to fuel moisture and correct a misplaced front in synthetic tests.} \textbf{a--c}, Head displacement from the 8\% moisture forecast at 12, 30 and 60\,s. Dashed line, prediction from the derivative at 8\% moisture. Blue circles, fresh coupled simulations. \textbf{d--i}, Fronts at 120, 180 and 300\,s from the first (d--f) and second (g--i) starting points: target fronts from the original WRF (black dashed), before correction (pink) and after one Gauss--Newton update (blue). Circles show synthetic observations at the fitted times. The 300-s forecast is not fitted. Scale bar, 200\,m. \textbf{j,k}, Signed radial position errors at 300\,s on 32 fixed rays: forecast minus target radius, with $0^\circ$ eastward and $90^\circ$ northward. \textbf{l}, Radial root-mean-square error (RMSE) of the 300-s forecast for both starts before correction and after wind-only, moisture-only or joint updates. \textbf{m--o}, Arrows connect the initial (pink squares) and updated (blue circles) inputs, which are the two wind offsets and the fuel moisture. Dashed lines mark the generating values.}
\label{fig:moisture}
\label{fig:inverse}
\end{figure}

One Gauss--Newton update uses the Jacobian to reduce the misfit. At 300\,s, a time not used in the fit, the root-mean-square error in front position falls from about 52 to 9\,m for the first starting point and from 28 to 7\,m for the second (Fig.~\ref{fig:inverse}d--i,l). Along fixed rays, we can also see where the corrected front still lies inside or outside the target. Most of the bias at the head of the fire disappears from both starts, and the remaining errors vary with direction (Fig.~\ref{fig:inverse}j,k).

From both starts, changing wind alone reduces the later front error more than changing moisture alone (Fig.~\ref{fig:inverse}l and Supplementary Table~\ref{S-tab:front-control-comparison}). From the second start, the wind-only update gives the smallest later error, about 3\,m. When we adjust wind and moisture together, the two starting points lead to different combinations that fit the front similarly well (Fig.~\ref{fig:inverse}m--o). Neither combination recovers the values that generated the fronts. The correction therefore improves the position of the front without identifying the winds and moisture behind it.

\clearpage
\subsection*{Coupled forecasts with neural component replacements}
\begin{figure}[p]
\centering
\includegraphics[width=.98\textwidth]{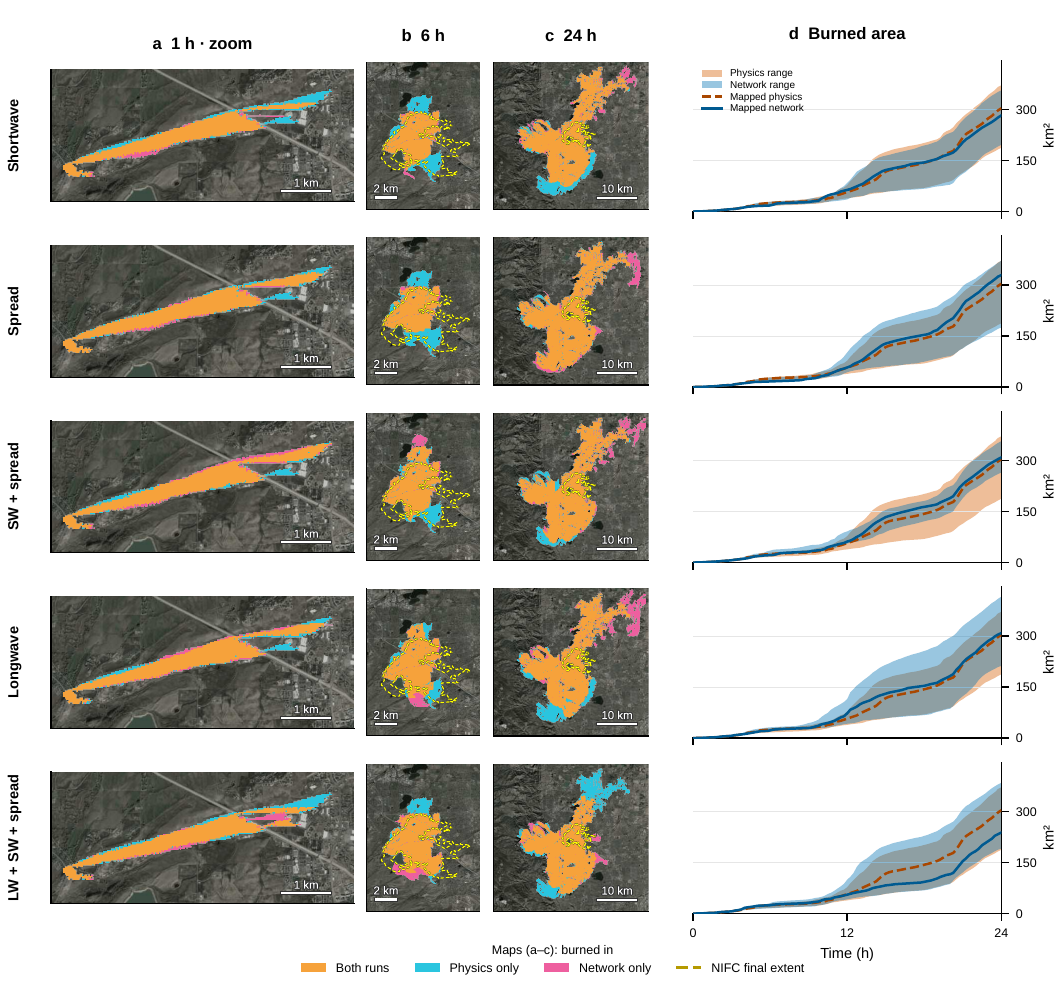}
\caption{\textbf{Neural components operate separately and together in a coupled fire forecast.} Rows show forecasts with the shortwave, spread, shortwave and spread, longwave and all three networks. \textbf{a}, One-hour footprints, zoomed. \textbf{b}, Six-hour footprints. \textbf{c}, Twenty-four-hour footprints. Map colors show burned overlap: orange, both runs; cyan, physics only; pink, network only. Yellow dashes mark the final extent mapped by the National Interagency Fire Center (NIFC) as geographic context. \textbf{d}, Burned-area histories. Orange (physics) and blue (network) shading show the ranges of ten 24-hour forecasts each. The physics members differ only by last-bit changes to the initial winds. Dashed orange and solid blue curves identify the physics and network forecasts mapped in a--c. The mapped physics run is the same physical forecast in every row. Basemap, Esri World Imagery.}
\label{fig:emulators}
\end{figure}

Neural networks can replace physical schemes inside the coupled forecast, and the forecast shows what each replacement changes. Each network returns the same quantities that a physical scheme passes to the rest of the model. We select radiation for its computational cost and fire spread for its physical importance and scope for improvement. As a demonstration and a starting point for training on observations, we train the networks on outputs of the physical schemes they replace. Five configurations replace shortwave (solar) radiation, spread, shortwave and spread together, longwave radiation, or all three, each with 24-hour forecasts (Fig.~\ref{fig:emulators}). Physical and network forecasts start from the same inputs and random-number state, then advance independently with atmospheric feedback and firebrands active.

We flip the last bit of the initial winds (one unit in the last place) to show how much fire growth can vary while all physical schemes stay fixed. These flips have no physical meaning, yet they grow into different fire histories as winds, heat release and individual firebrand events interact. We therefore compare each network forecast with an ensemble of perturbed physical forecasts as well as the unperturbed one, which puts differences in burned area and fire position on the scale of the model's numerical sensitivity.

The shortwave network passes sunlight down one atmospheric layer at a time. The longwave network predicts upward and downward fluxes, which determine how much each layer heats or cools. We also trained the longwave network on selected columns from the Marshall simulation, so this forecast reuses part of its training case. The spread network learns rate factors that depend on fuel properties, wind and slope. Fuel and moisture preprocessing and the direction of spread remain part of the physical calculation. The network also uses a different implementation of the fire-front calculation, so the comparison includes the effects of both changes. Methods and Supplementary Section~\ref{S-sec:independent-emulators} give the training details.

Figure~\ref{fig:emulators} follows the burned footprints from the first hour onward and shows their area histories alongside the perturbed physical forecasts. We measure each replacement, alone and combined, through burned area, footprint overlap on the fire grid and atmospheric fields, at the same forecast times and within pairs of network and physical forecasts that share a starting state. The full histories are essential, because area differences change sign and size as the fire evolves and similar totals can accompany different front positions.

Changing the last bit of the initial winds gives a 24-hour burned area of 187.5--371.5\,km$^2$ across ten physical forecasts. We ran each network configuration as a ten-member ensemble in the same way, and all five overlap this range (Fig.~\ref{fig:emulators}d). Their mean burned areas are higher than the physical mean by 2.9\% for shortwave, 4.0\% for spread, 16.2\% for shortwave and spread together, 26.1\% for longwave and 9.3\% for all three. The paired histories show how burned-area differences develop over time (Supplementary Fig.~\ref{S-fig:emulator-ensemble-paired}).

Similar burned areas can hide different footprints. The mapped longwave forecast burns only 1.6\% more area than the mapped physical forecast, yet the two agree on just 71\% of the area burned by either. For scale, we ran four additional perturbed physical forecasts, separate from the ten-member ensemble, and their overlap with the same physical forecast is 65--78\%.

The atmosphere differs more than the burned area. In the single shortwave, longwave and all-three forecasts, the differences in wind, surface temperature and surface heat and radiation fluxes all exceed the corresponding ranges from those four physical comparisons (Supplementary Table~\ref{S-tab:emulator-twenty-four-hour}). Supplementary Section~\ref{S-sec:ysu-review} reports a further experiment, in which a less accurate network replaces part of the boundary-layer scheme.

\section*{Discussion}
An established physical model can move into a machine-learning framework without changing its calculation. The standard must be exact because the calculation is sensitive. A firebrand that lands one grid cell farther downwind can start a fire in a different place. In the Marshall simulation, the corrected Fortran code and WRF-Speed agree on the winds, fire growth and firebrand population at every checkpoint through twelve hours. Both carry the fire beyond its observed final perimeter, so we can rule out the translation as the source of their shared excess spread. We also used the established calculation as a numerical standard\cite{baker2015cect} for checking the code written by coding agents. The radiation routine passed its own tests even when the surrounding code passed too little heating to the atmosphere. Comparing the assembled model with the Fortran code exposed the error.

Two different combinations of wind and fuel moisture produced close fits to the same fire front. In the idealized experiment, a single adjustment to winds and fuel moisture improves the front at a later time withheld from the fit. Yet the two starting points lead to different winds and moisture, and both corrected forecasts fit the later front well. A perimeter can therefore help correct the forecast while leaving the winds and moisture that produced the front uncertain. This distinction matters for efforts to estimate winds and moisture from observed fire fronts\cite{mandel2012perimeter,rochoux2015state}.

Networks have been coupled to atmospheric solvers and used together for moist physics and radiation\cite{kochkov2024neuralgcm,wang2022neuralphysics}. In a coupled fire forecast we can also follow the effects of individual and combined replacements all the way to the fire perimeter. The longwave network illustrates the value of doing so. Its forecasts overlap the range of total burned areas obtained by perturbing the physical model, yet its ensemble mean is a quarter higher. An individual forecast can also put a similar burned area in different places. The perturbations give each comparison a scale set by the physical model's own sensitivity.

The exact comparison covers one nested fire case. The timings and neural forecasts use the fast route. The idealized derivatives pass their checks, while the Marshall result depends on the wind perturbation being tested. An update from an observed fire perimeter would also need to account for errors in both the observations and the model. We trained the networks to reproduce existing schemes. Their effect on accuracy against observed fires remains untested.

The Rothermel formulation of fire spread dates to 1972 and describes quasi-steady surface-fire spread\cite{rothermel1972}. A network trained on observed fires could take its place and be tested inside the full forecast, as the spread network is here. Other communities rewriting established models can use their existing calculation in the same way, both to recognize errors already present and to measure what changes when a new scheme takes its place. The tested physics of an established model and the tools of machine learning can thus be had together.

\section*{Methods}

We first describe the model configuration that we rewrote (Code base and scope). To make the work of the coding agents checkable, we add short programs that print each Fortran routine's inputs, intermediate values and outputs to the last digit, so each check is against the compiled WRF code itself, not a hand-written imitation (Reference build). For the exact comparison we use double precision and evaluate functions such as exponentials and logarithms with the same mathematical library (glibc) in both codes\cite{goldberg1991floating} (WRF-Speed arithmetic). We translate each physical scheme in five checked stages (Translation protocol). We treat any difference from the Fortran code, however small, as a bug, and we find its source by giving each routine identical inputs and locating the first routine that disagrees (Diagnosing numerical differences). We then run the complete Marshall Fire configuration in both codes and compare their saved states every two simulated minutes (Coupled comparison). With this agreement established, we time the forecast and its derivatives on GPUs (Timing). We test the derivatives against finite differences (Gradient checks), and we use them to correct a misplaced fire front (Front-correction experiment) and to predict the response of the front to fuel moisture (Fuel moisture-response experiment). We also describe how burned area and the other plotted quantities are calculated (Diagnostic calculations and figure preparation). Finally, we train neural networks to replace physical schemes and run them inside the coupled forecast (Neural component training and coupled evaluation).

\subsection*{Code base and scope}
WRF holds about 800,000 lines of Fortran in its dynamics, physics and framework source. We rewrote selected WRF configurations in PyTorch and checked the rewrite against the Fortran code compiled in double precision. That Fortran code includes documented corrections (Reference build). Supplementary Table~\ref{S-tab:inventory} lists the reconstructed components, what each computes and the size of its Fortran source, about 131,000 lines in total. Supplementary Methods, Section~\ref{S-sec:execution}, describes the implementation and execution configurations.

The main comparison uses the configuration of the Marshall Fire simulation\cite{frediani2025spotting}. Its dynamical core is WRF-ARW, which advances each timestep in three Runge--Kutta stages and treats sound waves with shorter steps inside each stage (acoustic substeps)\cite{skamarock2019wrf,wicker2002rk3}. The vertical coordinate follows the terrain near the ground and flattens with height (the hybrid mass coordinate\cite{park2019hybrid}). A 1-km outer domain (the parent, $163\times159\times44$ grid cells, 3-s step) supplies boundary values to a 111-m inner domain (the nest, $360\times342\times44$ cells). The nest takes nine 1/3-s steps for each parent step, which we call nest substeps. The physics are the YSU boundary layer\cite{hong2006ysu} (parent only), the revised MM5 surface layer\cite{jimenez2012sfclay}, the Noah land surface\cite{chen2001noah}, WSM6 microphysics\cite{hong2006wsm6}, RRTMG longwave\cite{iacono2008rrtmg} and Dudhia shortwave\cite{dudhia1989} radiation. The fire model is WRF-Fire\cite{coen2013wrffire,mandel2011sfire}. It tracks the fire front with a level-set function, a field that is negative where the fire has burned and zero on the front. The front advances at the Rothermel spread rate\cite{rothermel1972}, a formula based on fuel properties, wind and slope, with the Anderson fuel categories\cite{anderson1982}. The fire grid is four times finer than the 111-m nest in each horizontal direction. The spotting scheme\cite{frediani2025spotting} follows individual firebrands (burning embers carried by the wind) as particles.

\subsection*{Reference build}
The Fortran code is WRF 4.5.1 compiled with GNU Fortran in double precision. We make the compiler call the standard one-value-at-a-time versions of functions such as the exponential and logarithm (scalar functions), because the faster versions that process several values at once can round differently. We compile the firebrand code without automatic loop vectorization, a compiler optimization that processes several loop iterations at once and can change the last bit of a result. To save intermediate values, we add output statements to copies of the Fortran code. Each modified copy must reproduce the unmodified calculation bit for bit before we use its output for comparison. The Fortran code used for comparison includes documented corrections (Supplementary Table~\ref{S-tab:corrections}). More than half concern the firebrand code, and the others concern the land-surface scheme, the fire-wind interpolation and restarts. The coupled Fortran code runs on 32 MPI ranks in a $4\times8$ layout.

\subsection*{WRF-Speed arithmetic}
Exact reproduction requires the same sequence of rounded operations\cite{goldberg1991floating}. We therefore keep the grouping of every expression as written in the Fortran code. We divide where the Fortran code divides and multiply by a reciprocal where it does so, because the two round differently. We also add the terms of each sum in the same order. Independent grid points and columns, and the separate forecasts of an ensemble, run in parallel, while calculations that depend on one another keep their order. These include vertical sweeps in which each level uses the result from the level before it, the Runge--Kutta stages and the acoustic substeps. We reproduce the precision with which WRF reads each configuration value. For example, WRF reads the damping coefficient in double precision from the namelist (its run configuration file). By contrast, it stores the Noah soil-layer thickness in single precision and then converts it to double precision.

We use two arithmetic routes. In the exact route, PyTorch executes each operation as it is called, without compilation. GPU versions of the exponential, logarithm and power functions can differ in the last bit from the CPU mathematical library (glibc) that the Fortran code uses. The exact route therefore evaluates these functions on the CPU with that library and returns the results to the GPU. Because PyTorch cannot differentiate through a call to an outside library, we supply the derivatives of these functions as formulas, so the exact route remains differentiable. We build lookup tables the same way, because small table differences persist into later nonlinear calculations. The exact route is the route we check against the Fortran code, and the twelve-hour Marshall comparison uses it. The fast route solves the same equations with standard GPU arithmetic and compiled kernels (small programs that run on the GPU). The forecast timings, the derivative timings and the forecasts with neural networks use the fast route.

\subsection*{Translation protocol}
AI coding agents wrote the WRF-Speed code, and we checked all of it against the compiled Fortran code. We set one rule for this work, namely that any difference from the Fortran code, however small, is a bug until a controlled test shows otherwise (see Diagnosing numerical differences). We never accepted a small remaining difference in order to make a comparison pass.

The agents rebuilt each physics scheme in five stages: (1) identifying the source code and settings that the configuration uses, (2) building compiled Fortran test cases that print inputs, intermediate values and outputs at full precision, (3) translating the calculation, (4) comparing intermediate quantities as well as final outputs and (5) saving the full-precision values as regression tests, which we rerun after every later change to the code. A stage had to pass before the next one began. We compared long routines section by section. We checked agreement at three levels: individual routines (e.g., one advection term, which transports a quantity with the wind, or one physics scheme), the coupling between components (i.e., that each value is passed in the right units and at the right time) and complete timesteps. The third level matters because a component can match on its own while the assembled model is wrong (see Results).

The AI coding agents used to build WRF-Speed were Claude (Anthropic)\cite{anthropic2026claude}, Codex (OpenAI)\cite{openai2026codex}, GLM 5.2 and 5.3\cite{glm2026glm5}, Kimi K2.7 and K3\cite{kimi2025k2,moonshot2026kimik3}, Grok 4.5\cite{xai2026grok45}, Gemini (Google)\cite{gemini2025gemini25}, Gemma (Google)\cite{gemma2025gemma3} and several DeepSeek\cite{deepseek2024v3,guo2025deepseekr1} and Qwen\cite{yang2025qwen3} models. The same models assisted with manuscript preparation under the authors' direction. The authors designed the verification procedures, directed the work and are responsible for the reported results.

\subsection*{Diagnosing numerical differences}

We investigate every nonzero difference with controlled tests of two kinds. In the first, we rebuild the Fortran code with one compiler setting changed and see whether the difference changes. One such setting is fused multiply--add contraction, in which the compiler merges a multiplication and an addition into a single operation that rounds once instead of twice. Another is the optimization level, for which we compare optimized and unoptimized builds. In the second, we supply WRF-Speed with intermediate values computed by the Fortran code, which shows whether a discrepancy originates in an earlier calculation.

To identify where a difference first appears within a timestep, we save the Fortran state over the entire domain after each routine of the timestep. We initialize WRF-Speed with inputs bitwise identical to the corresponding Fortran inputs and advance it one routine at a time. Comparing the outputs after each routine identifies the first routine at which the two codes diverge.

\subsection*{Coupled comparison}
WRF-Speed advances the coupled simulation in double precision on one NVIDIA L40 GPU using the exact route. The Fortran code and WRF-Speed each advance from their own state, with the same initial and lateral boundary data. We compare saved states every two simulated minutes, at 365 checkpoints. At each checkpoint we compare eight atmospheric fields per domain: three velocity components, potential temperature, geopotential, column mass, water vapor and pressure. For the fire, the comparison includes the fire level-set function, ignition time, fuel fraction, burned fraction, heat and moisture fluxes, the fuel burn-time constant and fuel category, and three firebrand counters kept in each fire-grid cell: firebrands generated, firebrands landed and spot ignitions.

At each checkpoint, we compare 68,064,683 values. Of these, we compare 62,154,923 floating-point values through their raw 64-bit patterns, so that even $+0$ and $-0$ count as different. The remaining 5,909,760 values are the fuel category and two of the three firebrand counters. We compare these numerically because their storage types differ between the two codes. The third counter has the same storage type in both codes and is in the bit-for-bit comparison.

In the coupled run, we compare the firebrands that are still in the air by their physical properties. The comparison ignores where each particle is stored, i.e., its position in the particle array and, in the Fortran code, which MPI rank holds it (a rank is one of the processes that share the work on a CPU cluster). Two firebrands with identical properties count as two, so the number of particles must also match.

\subsection*{Timing}
We time two calculations: a forecast step, and a forecast step followed by the backward pass that computes derivatives. For the forecast, one sample is one 3-s step of the 1-km domain together with the nine nest substeps it contains. Both codes run the same 100 steps of the Marshall simulation (03:01:06--03:06:06 UTC on 31 December 2021), of which five include a radiation update. The single-precision Fortran build is the executable used for the production Marshall Fire simulations\cite{frediani2025spotting}. The double-precision build uses the same GNU compiler configuration of WRF with its default optimization settings. It is a separate executable from the one used for the bit-for-bit comparison. Both run on one Derecho node with 128 MPI ranks. WRF-Speed runs on one A100 GPU and first advances 20 untimed steps (warm-up), so that one-time setup costs are excluded. We switch off output files in both codes. Each code starts from a model state saved by its own earlier run (a restart state), so the two cover the same simulated period but do not start from identical states.

To reduce the time the CPU spends sending work to the GPU, WRF-Speed records the GPU work of a timestep once and replays that recording on later timesteps (a CUDA graph, Supplementary Section~\ref{S-sec:execution}). Timesteps with a radiation update perform additional calculations that are not in the recording, so they run in the normal way. We made four runs of 100 steps each to test one optimization, and we report the mean time per step over the two runs that use it (Supplementary Section~\ref{S-sec:execution}). The model state at the end of each run is identical bit for bit to that of an earlier WRF-Speed run made without replay. Supplementary Table~\ref{S-tab:cost} lists the timings.

For the derivative timings, the 111-m benchmark advances nine nest substeps (3 simulated seconds) and then computes the derivative of one output with respect to two inputs. The output is the mean over the nest of $t_m$, the temperature variable that WRF advances (the moist potential temperature minus 300\,K). The inputs are uniform offsets added to the initial wind along the grid's $x$ direction and to that temperature variable. We compute the forcing from the 1-km domain and the radiative heating rates beforehand and hold them fixed. We repeat the calculation four times on the A100 and eight times on the GB200.

The reported times are wall-clock times, read after the GPU has finished its work. They exclude compilation and the one-time recording. On every repeat, the model state, the output and the derivatives must equal those of the same calculation run without compilation. This check confirms that the faster execution computes the same result. It does not test the derivatives themselves, which is the purpose of the gradient checks below. Supplementary Section~\ref{S-sec:execution} gives the individual samples and the memory use.

\subsection*{Gradient checks}
We test derivatives in three settings: one step of the 1-km Marshall domain, the fire model alone on idealized terrain and one substep of the 111-m Marshall nest. In each, we compare the derivative from automatic differentiation with a finite difference, obtained by rerunning the model with a slightly changed input. Automatic differentiation is the PyTorch facility that records the operations of a calculation and applies the chain rule to them.

For the 1-km domain, we evaluate derivatives of $J$, the mean over the domain of $t_m$ after one model timestep. Reverse-mode differentiation works backward from the output and gives the derivative of $J$ with respect to every input in one pass. Forward-mode differentiation carries one chosen input perturbation forward through the model. We compare the reverse-mode derivative in the uncompiled PyTorch implementation with central finite differences, calculated from model runs with equal positive and negative perturbations of size $h$. We use $h$ from $10^{-1}$ to $10^{-9}$ for three perturbation patterns: a uniform change in the grid-$x$ wind $U$, a uniform change in $t_m$ and a smooth random field.

We assess the derivatives using three criteria: (1) a relative difference from central finite differences of at most $10^{-6}$ at three consecutive perturbation sizes, (2) a Taylor remainder slope of $2.0\pm0.1$, and (3) agreement between forward- and reverse-mode directional derivatives to within $10^{-12}$. The Taylor criterion tests whether the error in a linear prediction of $J$ decreases approximately with the square of the perturbation size. The third criterion is a dot-product test. Forward mode gives the change in $J$ for a chosen input perturbation, and the reverse-mode derivative multiplied by the same perturbation must give the same number.

The dot-product test meets its criterion, with forward- and reverse-mode directional derivatives agreeing to within $1.1 \times 10 ^{-15}$ in four directions. By contrast, the finite-difference comparisons do not meet their criterion when model processes can switch on or off in response to the perturbations. For example, the WSM6 microphysics scheme computes deposition and sedimentation only where the cloud-ice and snow amounts are positive ($q_i>0$, $q_s>0$), and a perturbation can change where that holds. The Taylor test does not meet its criterion for the same reason.

We repeat the derivative checks while forcing conditional statements to retain the outcomes selected in the unperturbed run. With branch selections fixed, the $t_m$ derivative agrees with finite differences at four consecutive perturbation sizes, with relative differences of $2.7 \times 10^{-10}$ to $2.5 \times10^{-7}$ for $h=10^{-6}$ to $10^{-9}$. The uniform-$U$ derivative ($-4.24\times10^{-5}$\,K\,(m\,s$^{-1}$)$^{-1}$) agrees within tolerance at two perturbation sizes before changes in $J$ become too small to resolve in floating-point arithmetic. The derivative for the smooth random pattern also agrees at two perturbation sizes. Its agreement is limited by switches in the YSU boundary-layer scheme that we could not hold fixed. Thus, these two comparisons do not meet the requirement for agreement at three consecutive perturbation sizes.

The fire-model derivative checks in Supplementary Fig.~\ref{S-fig:fd} use WRF-Fire with prescribed uniform winds and no atmospheric integration. The simulations use double precision on CPUs, a $68 \times 68$ fire grid with 75-m spacing, and a 1-s timestep. We compare a 1,200-m-high hill with a radius of 2,250\,m against a flat-terrain control, using a 300-m ignition disk.

We evaluate the heat released by a fixed time in two regions: a disk of 750-m radius around the hill crest and the area at least 750\,m downwind of the ignition point. We do this at five operating points, each a combination of wind speed, wind direction and fuel moisture (e.g., 5\,m\,s$^{-1}$ upslope at 6\% moisture). Each derivative, for one operating point, terrain, input and region, is one entry. This gives 20 downstream entries and 18 crest entries, because on flat terrain the cross-slope case releases no heat in the crest disk. The region does not move when the inputs are perturbed. Upslope winds are directed along $+x$, and cross-slope winds along $+y$. We calculate derivatives with respect to uniform wind speed and dead fuel moisture.

We compare automatic derivatives, $D_{\rm AD}$, with central finite-difference derivatives, $D_{\rm FD}$, using 13 perturbation sizes from $10^{-9}$ to $10^{-3}$. A derivative meets the acceptance criterion when three consecutive perturbation sizes satisfy \[|D_{\rm AD}-D_{\rm FD}| \leq10 ^{-4} + 10^{-3}|D_{\rm FD}|.\] The absolute tolerance has units of GJ per m\,s$^{-1}$ for wind derivatives and GJ per kg\,kg$^{-1}$ for moisture derivatives. Here $Q$ is the heat release in the selected region, $x$ is the input (wind speed or fuel moisture) and $h$ is the perturbation size. When $|Q(x+h)-Q(x-h)|<10^{-12}|Q|$, the change in $Q$ is too small to resolve in floating-point arithmetic, and we do not count the comparison as a pass or a fail.

All 38 entries meet the criterion, although individual perturbation sizes can fail. A change in which fire-grid cells are burning does not explain these failures. The number of burning cells at each step is unchanged in 47 failed comparisons and changes in 55 successful ones. We fixed the test procedure before performing the 530 forward simulations.

For the coupled 111-m derivative check, we start the Marshall nest from its state 144 minutes into the simulation and perturb the grid-$x$ wind on the lowest three model levels over a fixed region. We calculate the reverse-mode derivative of a projected burned area after one 1/3-s substep, with atmospheric physics, fire spread, heat feedback and spotting active. We compare the derivative with central differences using wind perturbations of 0.01, 0.03, 0.1, 0.3 and 1\,m\,s$^{-1}$. The tolerance is $10^{-5}$\,acre\,(m\,s$^{-1}$)$^{-1}+10^{-3}|D_{\rm FD}|$. None of these increments meets the tolerance criterion. The recorded counts for burned area, level-set function, fuel and spot ignition do not change, but the largest perturbation adds two active firebrands.

The discrepancies have two sources. First, the fire level-set scheme (fifth-order weighted essentially non-oscillatory, WENO5) chooses which neighboring cells to use according to the direction in which the front moves, and a perturbation can change that choice. Second, the code limits the wind speed, so a perturbation has no effect where the limit applies. The WENO5 calculation matches the Fortran code. When we hold the conditional choices of the code (the places where it selects between alternatives) as they were in the unperturbed run, the derivative meets the tolerance at 0.01, 0.03, 0.1 and 0.3\,m\,s$^{-1}$ (absolute errors $3.3\times10^{-9}$ to $2.9\times10^{-6}$). With conditional choices allowed to change, the derivatives meet the tolerances for perturbations of 0.005\,m\,s$^{-1}$ and below.

\subsection*{Front-correction experiment}
The idealized coupled experiments use a flat 2-km domain with $40\times40\times50$ cells at 50-m spacing, a 12.5-m fire grid, a 0.25-s timestep, Anderson category 3 fuel (tall grass) and three-dimensional Smagorinsky turbulent mixing\cite{smagorinsky1963general}. Radiation, microphysics, land surface, boundary layer and spotting are off, while fire--atmosphere coupling is on. For the front-correction experiment, we generate synthetic fire fronts with the unmodified Fortran code of WRF~4.8 at a 0.125-s step. We use a different WRF version and half the timestep so that the observations do not come from the configuration that we correct. The observations are the distances to the front along 16 fixed rays at each of two times, 120 and 180\,s. We add noise with a standard deviation of 6.25\,m and a correlation length of 50\,m along the front. These observations constrain two initial wind offsets and initial fuel moisture.

Three forward-mode integrations, one for each input, provide a $32\times3$ Jacobian at each starting point. This matrix contains the derivatives of the 32 front positions with respect to the three inputs. The derivatives agree with central finite differences at three consecutive perturbation sizes, for every ray at both times. The gradient of the misfit between forecast and observed fronts, computed in reverse mode, also passes finite-difference checks. A Gauss--Newton update uses the Jacobian to estimate the change in the inputs that most reduces the misfit\cite{ledimet1986variational,rochoux2014parameter}. We fixed the update rule, its maximum size, the bounds on the inputs and the line search (Supplementary Section~\ref{S-sec:inverse-protocol}) before running any corrected forecast.

\subsection*{Fuel moisture-response experiment}
The moisture-response experiment uses the same flat 2-km geometry and 0.25-s atmospheric timestep as the front-correction experiment, with uniform initial grid-$x$ wind of approximately 3\,m\,s$^{-1}$ and a line ignition during seconds 0--5. We vary the initial fuel moisture uniformly on fueled cells and hold the dry fuel load fixed at $D=0.674/1.06$\,kg\,m$^{-2}$. To preserve the dry fuel load, we set the fuel loading input to $L=D(1+m)$, where $m$ is fuel moisture in kg\,kg$^{-1}$.

We locate the fire head by linear interpolation of the zero contour of the level set along five fixed transects at $y=843.75$, 918.75, 993.75, 1068.75 and 1143.75\,m. Their average defines the mean head position, $X(t)$. We calculate the mean spread rate between 9\,s and time $T$ as $[X(T)-X(9)]/(T-9)$. When differentiating this expression with respect to fuel moisture, we include the moisture dependence of both $X(T)$ and $X(9)$.

We check the derivatives of head position and mean spread rate against central finite differences using a combined absolute and relative tolerance. The relative tolerance is $10^{-3}$. The absolute tolerance is $10^{-3}$\,m per kg\,kg$^{-1}$ for head position and $10^{-5}$\,(m\,s$^{-1}$) per kg\,kg$^{-1}$ for spread rate. At 12 and 30\,s, both derivatives meet the tolerances at three consecutive moisture perturbations of $10^{-5}$, $3\times10^{-6}$ and $10^{-6}$\,kg\,kg$^{-1}$. At 60\,s the corresponding perturbations are $3\times10^{-6}$, $10^{-6}$ and $3\times10^{-7}$. We then compare new simulations at 6\%, 7\%, 9\% and 10\% moisture with the prediction from the derivative computed at 8\% moisture. Before doing so, we repeat the 8\% run and confirm that it reproduces the saved run exactly.

\subsection*{Diagnostic calculations and figure preparation}

We calculate burned area by integrating the model's burned fraction over the fire grid. For comparisons between physical and neural forecasts, we calculate the intersection over union, given by the area burned in both forecasts divided by the area burned in either forecast. This calculation uses binary burned-area masks defined by $\mathrm{LFN}<0$, where $\mathrm{LFN}$ is the fire level-set function. The masks use the original fire-grid coordinates and omit the extra, unused storage row and column. Fire-front correction figures use the saved level-set fields from the corresponding simulations without smoothing or resampling.

Timing figures show wall-clock times read after the GPU has finished its work and, separately, the times of the forward and backward parts measured on the GPU itself. We retain all timing samples, including one A100 repeat that took 17.86\,s against 13.73 to 13.74\,s for the other three. Supplementary Section~\ref{S-sec:figure-construction} gives the fire-grid coordinates, the two definitions of the burned footprint and notes on the figures.

\subsection*{Neural component training and coupled evaluation}
We train three neural networks, one each for shortwave radiation, longwave radiation and the fire spread rate. The training targets are the outputs of the physical schemes that the networks replace. We then run the networks inside the coupled model and measure how the simulated atmosphere and fire change.

Radiation training uses generated atmospheric profiles and columns from simulations of the 2020 Cameron Peak Fire in northern Colorado. We chose a different fire so that the Marshall forecasts test the radiation networks outside their main training case. The longwave training data include clear-sky profiles at low temperatures and cloudy profiles. We separate the Cameron Peak columns by date and spatial group into training columns and development columns, which we hold out to check the fit. We scale the inputs and outputs using means and standard deviations estimated from the training columns alone. The longwave network is the exception to the independent test. We train it further on outputs of the physical scheme for selected Marshall columns, again with separate training and development columns, so its Marshall forecast reuses part of its training case.

The longwave network uses three hidden layers with 512 units each and SiLU activation functions. The shortwave network applies the same two 32-unit SiLU layers at every vertical level and predicts the fractions of incoming radiation that are absorbed, scattered or transmitted, which sum to one. A separate fire spread network uses two hidden layers with 64 units each and tanh activation functions. The output of the spread network is the sum of a nonlinear prediction from the two hidden layers and a direct weighted sum of the inputs. The spread network has 4,757 trainable parameters and learns the rate factors by which fuel properties, wind and slope change the spread rate. The physical scheme still prepares the fuel and moisture inputs, sets the direction of spread, limits the spread rate and handles chaparral (shrub) fuel, which uses a separate formula. Supplementary Section~\ref{S-sec:independent-emulators} specifies the inputs, outputs, retained calculations, development datasets and errors.

Matched physical and neural forecasts start from the same initial conditions and stored random-generator state. Each forecast then advances independently from its own checkpoint. We compare fire geometry on the model grid, integrated burned fraction and atmospheric differences to quantify the accumulated coupled effects. The spread network runs through the uncombined version of the level-set code, while the physical scheme uses a faster kernel that combines several of its operations. Thus, these comparisons include the effects of both the network and its execution method.

Additional forecasts quantify sensitivity to small numerical perturbations: ten for the physical model, ten for each of the five network configurations and four further physical forecasts used for the footprint and atmospheric comparisons. Each perturbed physical forecast keeps the same physical schemes, forcing and random-number state as the unperturbed forecast, and changes the initial winds by one unit in the last place. These forecasts provide a scale for interpreting differences caused by the networks by showing how much the physical model changes in response to rounding-scale perturbations. We compare burned area, fire geometry and atmospheric fields at common valid times (Supplementary Sections~\ref{S-sec:emulator-coupled-scores} and~\ref{S-sec:nn-numerical-sensitivity}). Supplementary Table~\ref{S-tab:cost} lists the hardware and arithmetic route of each run. Supplementary Table~\ref{S-tab:evidence-scope} lists the configuration and comparison behind each main result.

\section*{Data availability}
The WRF reference source is available at \url{https://github.com/wrf-model/WRF}. Frediani et al.\cite{frediani2025spotting} describe the Marshall case and firebrand parameterization. Their source is available at \url{https://github.com/NCAR/WRF-FirebrandSpotting}. The initial and boundary files, restart states, configuration and the data behind the figures and tables are available to editors and referees during peer review. We will archive them on Zenodo with a DOI on publication.

\section*{Code availability}
WRF-Speed (wrf-torch, version 0.1.0) is developed at \url{https://github.com/NCAR/WRF-torch}. The source code, tests, corrections to the Fortran code and trained networks are available to editors and referees during peer review. On publication, we will release them under the Apache-2.0 license and deposit them on Zenodo with a DOI, together with the software versions, inputs and procedures required to reproduce the results.

\section*{Acknowledgements}
This material is based upon work supported by the NSF National Center for Atmospheric Research, which is a major facility sponsored by the U.S. National Science Foundation under Cooperative Agreement No.\ 1852977. We acknowledge high-performance computing support from Derecho\cite{derecho2023} and Casper (\url{https://ncar.pub/casper}) provided by the Computational and Information Systems Laboratory, NCAR, and sponsored by the National Science Foundation. We also acknowledge the Texas Advanced Computing Center (TACC) at The University of Texas at Austin for providing access to the Horizon system that allowed us to benchmark performance on NVIDIA GB200 GPUs. 

\section*{Author contributions}
JSS led the research, software development, numerical experiments and manuscript preparation. IL and JD contributed to model and software design and provided code examples. MF provided WRF-Fire details on the Marshall fire simulation and its Fortran oracle, and helped prepare the manuscript. PJ helped design the differentiability examples. All authors revised the manuscript.

\section*{Competing interests}
The authors declare no competing interests.

\clearpage
\setcounter{section}{0}
\setcounter{equation}{0}
\renewcommand{\thefigure}{S\arabic{figure}}
\renewcommand{\thetable}{S\arabic{table}}
\renewcommand{\theequation}{S\arabic{equation}}

\begin{center}
{\large\textbf{Supplementary Information}}\\[4pt]
An exact and differentiable rewrite of a coupled weather and fire model
\end{center}

\captionsetup[table]{name={Supplementary Table}}
\captionsetup[figure]{name={Supplementary Fig.}}
\setcounter{table}{0}
\setcounter{figure}{0}

\renewcommand{\thesection}{S\arabic{section}}
\renewcommand{\thesubsection}{\thesection.\arabic{subsection}}
\tableofcontents
\bigskip
\section{Execution of the rewritten model}
\label{S-sec:execution}
WRF-Speed runs in two ways. The exact route executes each PyTorch operation as it is called and uses the same mathematical library as the Fortran code. We claim bit-for-bit agreement only for this route. The fast route uses standard GPU arithmetic and compiled kernels (small programs that run on the GPU), and it is the route we time and differentiate on large domains. Throughout, the parent is the 1-km outer domain and the nest is the 111-m inner domain that carries the fire. This section describes how the fast route is built and how we measured its costs. We check each change to the fast route against the fast-route calculation without that change, not against the exact route or the Fortran code. Supplementary Table~\ref{S-tab:evidence-scope} lists the calculation behind each result in the paper, what it was compared with and its limits.

\begin{table}[p]
\centering\footnotesize
\caption{\textbf{What each reported result was calculated with and compared against.} The exact trajectory, the GPU timings, the derivative checks and the neural forecasts are separate calculations. All GPU timings, derivative benchmarks and neural forecasts use the fast route, so they are compared with fast-route or uncompiled PyTorch calculations and not with the Fortran code.}
\label{S-tab:evidence-scope}
\begin{tabular}{L{2.5cm}L{4.9cm}L{4.4cm}L{4.5cm}}
\toprule
Result & Calculation and configuration & Compared with & Limits\\
\midrule
Bit-for-bit agreement & Nested Marshall calculation in double precision, WRF-Speed exact route on one GPU & Corrected WRF~4.5.1 Fortran code on 32 MPI ranks at 365 checkpoints over twelve hours. Each code advances from its own state & Applies to the compared fields and to this build of the Fortran code\\
\addlinespace
Forward speed & Fast route on one A100 over 100 parent steps (95 without and five with a radiation update). Four runs, with a mean of 1.166\,s for the two with grouped advection & Wall time of the Fortran code on 128 CPU cores over the same model-time window & Different hardware and separately produced starting states. Startup, compilation and output are excluded\\
\addlinespace
Forward-and-backward speed & Nine 111-m nest substeps with parent forcing and radiation tendencies held fixed. Four A100 repeats and eight GB200 repeats per configuration. One compiled step on the 1-km domain & State, loss and gradients of the uncompiled PyTorch calculation from the same code version, reproduced at every repeat & Shows that faster execution preserves the calculation. It is not a finite-difference test of a Marshall forecast sensitivity. The A100 and GB200 runs differ in code, build and checkpoint placement\\
\addlinespace
Derivative checks, idealized & Fire model with prescribed winds at five operating points, the coupled moisture response and the front-position Jacobian & Central finite differences at three consecutive increments & The moisture and front tests pass. In the fire-model test not every comparison meets the criterion (Methods)\\
\addlinespace
Derivative checks, Marshall scale & One substep of the 111-m nest with a wind perturbation, and one step of the 1-km domain in three directions & Central finite differences & At 111\,m all five prespecified unmodified increments fail. Four increments pass with branch choices held fixed, and smaller unmodified increments at or below 0.005\,m\,s$^{-1}$ also pass. At 1\,km only one direction meets the three-increment rule, with branch choices held fixed\\
\addlinespace
Front correction & Idealized coupled domain, with synthetic fronts from the original WRF~4.8 and two starting points & Front position at the held-out time of 300\,s. RMSE changes from 51.57 to 8.57\,m and from 28.08 to 6.68\,m & The true inputs are not recovered. No validation against an observed fire\\
\addlinespace
Single neural forecasts & Five 24-hour Marshall forecasts with shortwave, spread, both, longwave or all three replaced by networks & The physical forecast in the same configuration, advancing on its own, with four perturbed physical forecasts as a scale & Single examples of the coupled response to these particular networks\\
\addlinespace
Ten-member ensembles & Ten 24-hour forecasts for the physical model and for each of the five network configurations, with last-bit changes to the initial winds & Mean and range of burned area in the physical ensemble & Burned area only. The networks were not tuned to match the physical ensemble\\
\addlinespace
Boundary-layer (YSU) networks & One 24-hour forecast with networks replacing parts of YSU, and ten six-hour pairs & The physical forecast at 24 hours and the physical ensemble at six hours & The 24-hour neural and physical forecasts use different code versions, so that comparison does not isolate the effect of the networks. The offline fit is weak\\
\bottomrule
\end{tabular}
\end{table}

\subsection{GPU kernels and CUDA graphs}
The rewrite first expresses each Fortran calculation as operations on whole arrays (tensors). Independent grid cells and columns are computed together, and steps that depend on one another keep their order. This form runs on a GPU, but slowly. Many array operations each launch a separate GPU kernel (a launch is one request from the CPU to run a kernel) and write intermediate results to memory for the next operation to read.

We reduce this cost in two ways. First, a compiler groups compatible array operations into GPU kernels, including the operations of the backward pass. Second, we wrote CUDA kernels by hand for selected routines, including advection (transport by the wind), the acoustic updates (the short substeps that handle sound waves) and parts of the physics. Each hand-written kernel keeps the arithmetic expressions and their order. It falls back to the original PyTorch code (or stops with an error) when its inputs do not have the expected form. For example, combined kernels for the WSM6 microphysics scheme avoid writing temporary arrays to memory, and the advection calls for the moisture species in the first two of the three stages of each timestep are grouped into one launch. Both versions use double precision. We check each such change against the fast route without the change. The comparison with the Fortran code uses the exact route.

A CUDA graph changes how the kernels are sent to the GPU. We record the GPU work for a timestep once and then replay it as a single launch, so the CPU no longer sends each kernel separately. For the forecast timings, one graph covers a parent step and its nine nest substeps. The forecast still advances step by step, because each replay starts from the state that the previous step produced. Steps that recompute radiation perform calculations that are not in the recorded sequence, so they run the normal way, without replay. This replay of an advancing forecast supports forward runs only, in double precision and on one GPU, with at most two domains and no feedback from the nest to its parent. Feedback from the fire to the atmosphere within the nest stays active. No replay was rejected in the four timing runs. The differentiation benchmark uses a different graph. It records a forward and a backward pass, and it is reset to the same initial state between repeats.

\subsection{Forward timing}
Methods describes the timing window. We add two details here. First, the Fortran timer for the parent domain already includes the nest, so we do not add the nest timer to it. Second, we wait for the GPU to finish before reading the wall clock. The timings exclude startup, compilation, graph recording and output, and they keep every measured step, including the slower radiation steps.

We made four timing runs to test one optimization, which sends the transport calculations for all the moisture variables (water vapor, cloud and precipitation) to the GPU together and not one variable at a time. The runs had this grouping off, on, on and off, in that order, and nothing else varied. The four individual run means are in the data archive described under Data availability. With grouped advection, the mean parent step takes 1.166\,s (1.118\,s without and 2.084\,s with a radiation update). Without grouping, the mean is 1.170\,s. Grouping therefore saves about 0.3\%. The radiation steps are 0.6\% slower in the grouped runs and remain in the mean. A separate profile shows 416 fewer GPU kernels with grouping. The saved states at the end of all four runs are bitwise identical to each other and to the earlier WRF-Speed run over this window.

\subsection{Differentiated execution}
The 111-m benchmark differentiates nine nest substeps. We add uniform offsets to the initial grid-$x$ wind and to the moist potential temperature, and we compute how the grid mean of $t_m$ after nine 1/3-s steps changes with each offset (Methods). The derivatives come from reverse-mode differentiation, the method known in machine learning as backpropagation. Each repeat starts again from the same initial state. Parent forcing and radiation tendencies (the heating rates that the radiation scheme adds to the model state) are held fixed. The separate 1-km benchmark compiles one step, its objective and its derivative together, and takes a median GPU time of 167.7\,ms (Supplementary Table~\ref{S-tab:cost}).

Reverse-mode differentiation needs intermediate states from the forward run. To limit memory, we keep intermediate arrays only at the boundaries between regions of the calculation and recompute the operations inside each region during the backward pass. On the A100 the arrays are kept in CPU memory, which takes 20.1\,GiB in addition to the GPU memory, and the transfers between CPU and GPU in both directions are part of the measured time. On the GB200 they stay on the GPU. We check all hand-written and compiled derivative code against PyTorch's automatic differentiation of the original code. We test separately whether the derivatives agree with finite differences (Methods, Gradient checks). The two GB200 configurations differ only in two execution options, and their states, losses and gradients agree bit for bit at every repeat. The A100 and GB200 runs differ in code version, software build and checkpoint placement as well as hardware, so their times are not a pure hardware comparison.

\subsection{Timing samples, memory and software}
The forward timings are mean wall times over 200 parent steps, 100 from each of the two runs with grouped advection. The 111-m forward-and-backward times are median wall times over four repeats on the A100 and eight repeats per configuration on the GB200. We read the wall clock only after the GPU has finished its work. The forward and backward phase times are medians timed on the GPU itself, and they exclude the reset and the checks between repeats. On the GB200, we measured the phase times in a separate run from the total wall times. All timings exclude startup, compilation, graph recording and warm-up. The forward window includes new radiation calculations, and the differentiation benchmarks do not.

The Fortran timings use WRF source revision \texttt{cde4a22a} on one Derecho node, which has 128 AMD processor cores, with one MPI rank per core. We built the double-precision executable with GNU Fortran 12.2.0 and Cray MPICH, using WRF's standard GNU configuration for distributed-memory runs. Real numbers are promoted to eight bytes (\texttt{-fdefault-real-8 -fdefault-double-8}), and the optimization flags are the defaults of that configuration (\texttt{-O2 -ftree-vectorize -funroll-loops}). The 128 ranks are arranged $8\times16$. The single-precision executable is the GNU build used for the production Marshall Fire simulations\cite{frediani2025spotting}.

The four A100 wall times are 13.73, 13.73, 13.74 and 17.86\,s. We keep all four and report the median, 13.73\,s. In the 17.86-s sample the GPU phases take 13.73\,s, so the extra time is not GPU computation.

Memory depends on the stage, because recording a graph needs more memory than replaying it. On the A100, the allocated peak is 62.5\,GiB over the whole run, 54.2\,GiB while the graph is recorded and near 15.0\,GiB during replay. For the 111-m rows, Supplementary Table~\ref{S-tab:cost} reports the peak over the whole run, which includes preparation and recording, not the smaller replay requirement. The A100 benchmark uses PyTorch 2.14.1 with CUDA 12.6, and the GB200 benchmark uses PyTorch 2.14.1 with CUDA 13.0. The individual timing samples, the memory at each stage on both GPUs and the software and execution settings are in the data archive described under Data availability.

\section{Derivative tests and front correction}
\label{S-sec:derivative-qualification}
\begin{figure}[!t]
\centering
\includegraphics[width=\textwidth]{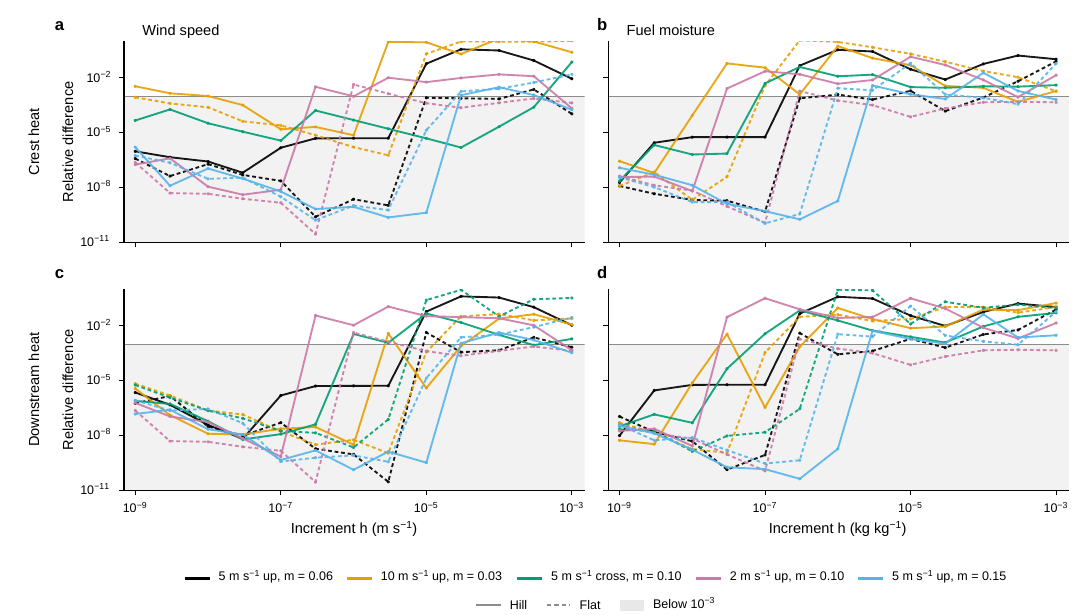}
\caption{\textbf{Automatic derivatives compared with finite differences in the fire model.} Relative difference against increment $h$ for crest-disk (a,b) and downstream (c,d) heat release, differentiated with respect to wind (a,c) or moisture (b,d). Colors mark five operating points, and solid and dashed lines show hill and flat terrain. The gray band marks the relative term ($10^{-3}$). The full acceptance criterion is given in Methods.}
\label{S-fig:fd}
\end{figure}

Methods describes how we test derivatives against finite differences, in which the model is rerun with a slightly changed input and the change in the output is divided by the change in the input. Supplementary Fig.~\ref{S-fig:fd} shows the fire-model test, which covers derivatives with respect to wind and fuel moisture at five operating points on hill and flat terrain. The crest disk is centered on the hill at (2,587.5, 2,587.5)\,m, and the ignition disk at (1,087.5, 2,587.5)\,m. Not every comparison meets the criterion, and Methods reports the ones that fail. The two Marshall-scale checks pass only in part. At 111\,m, none of the five prespecified increments meets the criterion with unmodified finite differences. With the branch choices of the unperturbed run held fixed, four of them do, and unmodified increments of 0.005\,m\,s$^{-1}$ and below also meet it. At 1\,km, unmodified finite differences do not meet the criterion, and with branch choices held fixed only one of the three directions meets the three-increment rule. The remainder of this section gives the procedure for the synthetic front correction. Supplementary Table~\ref{S-tab:moisture-response} lists the displacements behind the moisture-response comparison in the main text (Fig.~\ref{fig:moisture}a--c).

\begin{table}[htbp]
\centering\small
\caption{\textbf{Predicted and simulated change in fire-front position with fuel moisture.} Changes in mean head position after 60\,s, relative to the forecast at 8\% fuel moisture. Predicted values come from the derivative computed in the single 8\% run, simulated values come from fresh coupled simulations at each moisture, and the error is the difference between the two as a percentage of the simulated change.}
\label{S-tab:moisture-response}
\begin{tabular}{rrrr}
\toprule
Fuel moisture (\%) & Predicted change (m) & Simulated change (m) & Error (\%)\\
\midrule
6 & 2.29 & 2.73 & 16.0\\
7 & 1.15 & 1.23 & 7.1\\
9 & $-$1.15 & $-$1.11 & 3.6\\
10 & $-$2.29 & $-$1.95 & 17.7\\
\bottomrule
\end{tabular}
\end{table}

\subsection{Synthetic front-correction protocol}
\label{S-sec:inverse-protocol}
The front correction adjusts three inputs (two wind offsets and fuel moisture) so that the forecast fire front matches synthetic observations of the front. This subsection gives the full procedure. The control vector is $c=(\Delta U,\Delta V,m)$, where the initial horizontal-wind offsets have units m\,s$^{-1}$ and $m$ is fuel moisture in kg\,kg$^{-1}$. Optimization uses $q=(c-c_0)/s$ elementwise, with $c_0=(0,0,0.09)$ and $s=(0.5,0.5,0.02)$ in the corresponding units. The objective is
\[
\mathcal{L}(q)=\frac{1}{64}\,[r(q)-y]^{\mathsf T}C^{-1}[r(q)-y],
\]
where $r$ contains 16 radial distances at 120\,s followed by 16 at 180\,s. Each is the distance along a fixed ray at which the ray first leaves the burned area. There is no penalty for moving away from the starting controls. The fixed covariance has independent time blocks and within each block
$C_{ij}=6.25^2[0.95\exp(-d_{ij}/50)+0.05\delta_{ij}]$\,m$^2$, where $d_{ij}$ is the shortest distance in meters between rays $i$ and $j$ measured along the front. We use the same synthetic noise realization for both starts.

With the position Jacobian $A=\partial r/\partial q$ (checked against finite differences, Methods), the update uses $g=A^{\mathsf T}C^{-1}(r-y)/32$, $H=A^{\mathsf T}C^{-1}A/32$ and $\delta q=-(H+\lambda I)^{-1}g$, with $\lambda=10^{-3}\max(\max\operatorname{diag}H,10^{-12})$. This is Gauss--Newton curvature, not an exact Hessian. The Euclidean step norm is capped at 3. Physical trial controls are projected into $\Delta U,\Delta V\in[-1.5,1.5]$\,m\,s$^{-1}$ and $m\in[0.04,0.14]$. Fresh forward calculations try step multipliers $1,1/2,1/4,1/8,1/16$ in order, accepting the first strict decrease of the actual objective. The stopping threshold, fixed in advance, is $\mathcal{L}\leq0.665$ after an accepted update, with at most three accepted updates per start. Both starts stop after one full step, and neither the cap on the step norm nor the bounds on the controls come into play.

The two starts are $(-0.75,0.50,0.09)$ and $(0.50,-0.50,0.06)$. Their accepted controls, respectively, are
\[
(0.09065,-0.25982,0.071547)\quad\text{and}\quad(-0.30470,0.02455,0.049112).
\]
Objectives decrease from 11.0223 to 0.6551 and from 3.7599 to 0.6482. The generating controls are $(0,0,0.07)$. The two corrections arrive at different controls, and neither equals the generating ones. Changes in wind and moisture can compensate for each other, so the fronts are matched without recovering the true inputs. The 300-s front is excluded from the objective, line search and stopping rule.

\begin{table}[htbp]
\centering
\caption{Front-position RMSE at 300\,s after one accepted correction. Errors in meters use 32 fixed rays against the noise-free generating forecast. The fitted observations are at 120 and 180\,s.}
\label{S-tab:front-control-comparison}
\begin{tabular}{lrrrr}
\hline
Starting point & Initial & Wind only & Moisture only & Joint\\
\hline
First & 51.57 & 11.16 & 31.79 & 8.57\\
Second & 28.08 & 3.15 & 14.54 & 6.68\\
\hline
\end{tabular}
\end{table}

\paragraph{Restricted-control comparisons.}
We recompute one update using either the two wind offsets or moisture alone, with the inactive controls fixed at their initial values. For the active controls $I$, the step is $\delta q_I=-(H_{II}+\lambda I)^{-1}g_I$, with the same $\lambda$, objective, observations, bounds and step rules as above. From the first start, the moisture-only update reaches the lower bound on moisture. Both updates from the second start accept the full step. Wind adjustment gives the smaller 300-s error from both starts (Supplementary Table~\ref{S-tab:front-control-comparison}), although moisture stays at its initial value. From the second start it also outperforms the joint update at 300\,s. The accepted controls and objectives of the restricted updates are in the data archive described under Data availability.

\section{Fire-grid arrays and figure notes}
\label{S-sec:fire-grid-registration}\label{S-sec:figure-construction}
WRF tracks the fire with a level-set function, which is negative where the fire has burned. It stores this function with one extra row and one extra column ($1441\times1369$ stored nodes for a $1440\times1368$ fire grid). The model never updates them, and a contour drawn through them would produce a false outline along the edge of the domain. We therefore omit this unused row and column and draw fire perimeters from the nodes that the model advances, at their original coordinates. Every contour is kept, including spot fires and unburned holes. Burned area is the sum over fire cells of the burned fraction times the cell area.

\begin{figure}[!ht]
\centering
\makebox[\textwidth][c]{\includegraphics{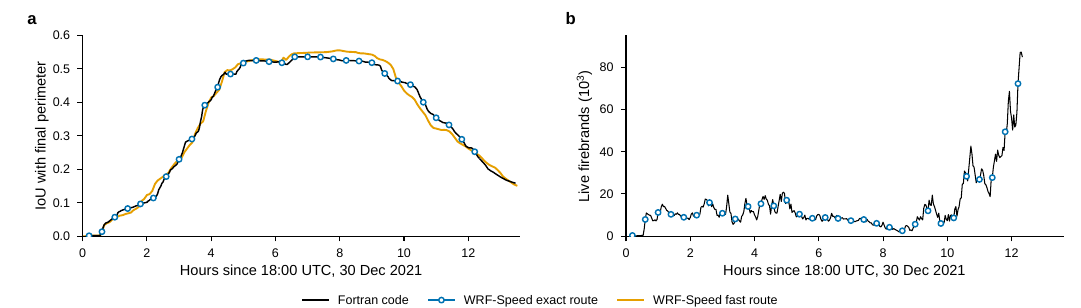}}
\caption{\textbf{Perimeter overlap and firebrands in the Marshall run.} \textbf{a}, Intersection over union with the observed final perimeter for the Fortran code (black), the WRF-Speed exact route (blue circles) and the fast-route run (orange). \textbf{b}, Live firebrands, which agree exactly between the Fortran code and the WRF-Speed exact route.}
\label{S-fig:edseries}
\end{figure}

We use two definitions of the burned footprint. The overlap with the observed perimeter (Fig.~\ref{fig:parity} and Supplementary Fig.~\ref{S-fig:edseries}) counts cells with a nonzero burned fraction. The comparisons between neural and physical forecasts count cells where the level-set function is negative. The observed perimeter is the final perimeter from the National Interagency Fire Center (NIFC/WFIGS, 6,079.90 acres). The Fortran code substantially overpredicts the observed final area, so matching it shows that the calculation is preserved, not that the forecast matches the observed fire.

The front-correction figure (Fig.~\ref{fig:inverse}) plots the saved level-set fields without smoothing or resampling. In Fig.~\ref{fig:capability}, wall times and phase times are medians of four A100 or eight GB200 repeats, and memory bars show the peak at each stage. Phase times are measured on the GPU. Both GB200 phase times, forward and backward, come from a separate run from the wall times. The Fortran forward costs pool two repeats of 100 parent steps, and the A100 forward cost pools the two grouped-advection runs over the same model-time window. In Supplementary Fig.~\ref{S-fig:fd}, the crest-disk heat release is zero for the flat cross-slope case, which is omitted.

\section*{Supplementary figures and tables}
\phantomsection\addcontentsline{toc}{section}{Supplementary figures and tables}

\begin{table}[ht]
\caption{\textbf{Components rewritten for the Marshall calculation.} We checked each component on its own against the Fortran code before running them together. The coupled comparison at 365 checkpoints is the test of the complete model. Line counts cover the WRF source files that contain the routines used. Whole files are counted, so the driver files include branches for schemes that this configuration does not select.}
\label{S-tab:inventory}
\footnotesize
\begin{tabular}{L{4.6cm}L{8.4cm}rr}
\toprule
Component & What it computes & Files & Fortran lines\\
\midrule
ARW dynamics & Winds, pressure and temperature of the atmosphere & 12 & 50,707\\
One-way nesting & Passes the forecast of the 1-km parent to the 111-m nest & 5 & 9,091\\
YSU, revised MM5 surface layer and Noah & Mixing near the ground, and heat and moisture exchange with the land & 4 & 13,378\\
WSM6, RRTMG longwave and Dudhia shortwave & Clouds and precipitation, and heating by infrared and solar radiation & 3 & 17,851\\
Physics drivers and initialization & Code that calls the schemes and applies their results & 6 & 27,119\\
Surface-fire level set & How the fire front spreads and how much heat it releases & 7 & 8,228\\
Firebrand spotting & Burning embers carried by the wind that start new fires & 1 & 4,516\\
\midrule
Total & & 38 & 130,890\\
\bottomrule
\end{tabular}
\end{table}

\begin{table}[ht]
\caption{\textbf{Recorded model execution costs.} Rows for the differentiable step use precomputed radiation tendencies, and the 111-m rows cover nine substeps with parent forcing held fixed. Forward-only values are mean wall times. Forward values of the differentiable step are medians timed on the GPU, the 111-m forward-and-backward values are median wall times, and the 1-km forward-and-backward value is a median timed on the GPU. Memory is the peak allocated over the whole run. The A100 111-m run also uses 20.1\,GiB of CPU memory for checkpoints. All rows time the physical model, and we report no speedup from the neural networks. n.m., not measured. Section~\ref{S-sec:execution} describes the routes and sampling.}
\label{S-tab:cost}
\small
\begin{tabular}{L{7.0cm}rrr}
\toprule
Calculation & Forward & Forward and backward & Memory\\
\midrule
\multicolumn{4}{l}{\emph{Differentiable step}}\\
111-m nest, nine substeps, A100 80\,GB & 2.68\,s & 13.73\,s & 62.5\,GiB\\
111-m nest, nine substeps, GB200, launch options & 1.36\,s & 5.76\,s & 114.1\,GiB\\
111-m nest, nine substeps, GB200, default launch & 1.50\,s & 6.39\,s & 114.1\,GiB\\
1-km domain, one step compiled as one program, A100 & 52.1\,ms & 167.7\,ms & n.m.\\
\midrule
\multicolumn{4}{l}{\emph{Nested parent step, forward only}}\\
Fortran double precision, one CPU node & 6.075\,s & -- & --\\
Fortran single precision (production), one CPU node & 2.769\,s & -- & --\\
WRF-Speed final configuration with CUDA graph replay, A100 80\,GB & 1.166\,s & -- & n.m.\\
\bottomrule
\end{tabular}
\end{table}

\begin{table}[ht]
\caption{\textbf{Corrections to the Fortran code used in the coupled comparison}, grouped by the calculation they affect. They address nondeterminism, undefined behavior, restart persistence, halo values at the boundaries between MPI ranks and one diagnostic, and some of them change outputs relative to the uncorrected source.}
\label{S-tab:corrections}
\small
\begin{tabular}{L{3.6cm}L{11.2cm}}
\toprule
Affected calculation & Corrections applied to the Fortran code\\
\midrule
Particle exchange & Initialize the receive mask on empty ranks, match MPI buffer width to eight-byte reals, and check incoming particle capacity.\\
Persistent particle state & Preserve live firebrands during restart initialization.\\
Fire--air exchange & Update roughness halos before fire-wind interpolation at the boundaries between MPI ranks.\\
Undefined outputs & Read the generation-cap diagnostic before deallocation, and assign canopy resistance on the glacial branch.\\
Maximum distance & Accumulate the maximum deposited distance in each cell rather than the last deposit.\\
Restart output scheduling & Preserve the repeating alarm period and the pending first history output time.\\
\bottomrule
\end{tabular}
\end{table}

\clearpage

\clearpage

\section{Radiation and spread networks}
\label{S-sec:independent-emulators}
We trained small neural networks to stand in for three physical schemes and ran each inside the full coupled Marshall forecast. The purpose is to show that a learned component can be inserted and measured, not to propose these networks for operational use. The networks replace the Dudhia shortwave radiation scheme\cite{dudhia1989} (heating by sunlight), the RRTMG longwave radiation scheme (heating and cooling by infrared radiation) and the Rothermel spread-rate formula\cite{rothermel1972} (how fast the fire front moves for a given fuel, wind and slope). Supplementary Table~\ref{S-tab:nn-offline} states what each network learns and which physical calculations are kept around it. We test five combinations: shortwave, spread, shortwave with spread, longwave, and all three. Each neural forecast and the physical forecast it is compared with start from the same Marshall initial state and use the same configuration, input files, grids, timesteps and random-number state. Each then advances on its own. All use the fast route in double precision, so these forecasts are separate from the bit-for-bit comparison with the Fortran code. The forecasts start at 18:00 UTC on 30 December 2021 from the initial conditions and lateral boundary forcing of the original Marshall case, so the six- and 24-hour results correspond to 00:00 and 18:00 UTC on 31 December. The coordinates and timing of the prescribed ignition are recorded in the data archive described under Data availability.

\subsection{Training data}
\label{S-sec:nn-training}
Radiation training uses atmospheric profiles from the 2020 Cameron Peak Fire in northern Colorado (13--15 August 2020), with development profiles from 6--7 September 2020, together with generated profiles. Development data are held out from training and used to select and evaluate the networks. We chose this different fire on purpose, so that the Marshall forecasts test the radiation networks outside their main training case.

Only the longwave network is also trained on columns from the physical Marshall forecast, taken at 4, 8, 12, 16 and 20 hours. These supply 16,640 columns from the parent and the nest, divided into 7,680 training and 8,960 development columns by saved time and by tile (a block of neighboring columns). All columns at 8 and 20 hours belong to development. Each training batch draws 52 Cameron Peak, 26 cold clear-sky, 25 cloudy and 25 Marshall columns. Inputs and targets are scaled with statistics from the original training data only, and the additional training penalizes departures from the earlier fit on the original data. The network takes the 44 model levels, the 45 level interfaces and surface values as inputs, and it returns fluxes and heating rates on the model levels. The Cameron Peak development profiles are all clear sky. Ozone and greenhouse-gas amounts are fixed as in the physical scheme, and cloud particle sizes are not inputs.

The shortwave fit uses 90 generated and nine Cameron Peak training families, and 100 generated and six Cameron Peak development families. A family is a group of related profiles that is kept together in either training or development. Only two Cameron Peak training columns and no Cameron Peak development columns contain cloud, so we report the generated-profile and Cameron Peak errors separately.

The spread network is trained on 32,768 evaluations of the physical spread-rate formula from 64 families of inputs, with 8,192 evaluations from 16 separate families for development. The training covers twelve of the Anderson fuel categories. Category 4 (chaparral) follows a separate formula in WRF-Fire, which we keep. The five inputs are transformed values of the packing ratio, the wind exponent and coefficient, the wind speed and the slope. The targets come from the physical formula, not from observed fire perimeters. Supplementary Table~\ref{S-tab:nn-training} lists the training settings.

\begin{table}[htbp]
\centering\small
\caption{\textbf{Training settings of the radiation and spread networks.} The values come from the saved training records of the networks used in the forecasts. The shortwave network was trained in two stages. The longwave network was trained in several stages, and the table lists the last stage before the Marshall columns were added and the fine-tuning with Marshall columns. For the longwave network, an epoch is 36 updates.}
\label{S-tab:nn-training}
\begin{tabular}{@{}L{3.3cm}L{2.5cm}L{2.0cm}rL{2.9cm}L{4.4cm}@{}}
\toprule
Network and stage & Optimizer & Learning rate & Batch & Training length & Selection\\
\midrule
Shortwave, first stage & Adam & $10^{-3}$ & 512 & 100 epochs & Lowest development error, averaged equally over generated and Cameron Peak data\\
Shortwave, second stage & Adam & $3\times10^{-4}$ & 512 & 300 further epochs & Same criterion\\
\addlinespace
Longwave, last stage before Marshall columns & AdamW, no weight decay & $10^{-4}$ & 128 & 43 epochs & Lowest development objective. Training stops after 15 epochs without improvement\\
Longwave, fine-tuning with Marshall columns & AdamW, no weight decay & $10^{-4}$, then $10^{-5}$ & 128 & 30 epochs, then 85 epochs & Lowest Marshall surface-flux RMSE among the epochs that meet the three conditions in Section~\ref{S-sec:nn-offline-errors}. Training stops after 30 epochs without improvement\\
\addlinespace
Spread & Adam, with a cosine learning-rate schedule & $2\times10^{-3}$, falling to $2\times10^{-5}$ & 2,048 & 15,000 updates & Lowest development error of the three log responses\\
\bottomrule
\end{tabular}
\end{table}

\subsection{Offline development errors}
\label{S-sec:nn-offline-errors}
Supplementary Table~\ref{S-tab:nn-offline} lists the errors of each network against the physical scheme it replaces, evaluated on its own outside the coupled model. All scores are on development data. The shortwave and spread networks follow their physical schemes closely, with $R^2$ (the fraction of variance explained, 1 for a perfect fit) above 0.99 for every quantity. The longwave network is the least accurate of the three.

\begin{table}[htbp]
\centering\small
\caption{\textbf{The three networks and their offline development errors.} For each network, the first row states what is learned, which physical calculations are kept and which constraints the output stage of the network enforces by construction. Parameter counts include the learned linear output layers. The errors compare each network with the physical scheme it replaces, on development data held out from training. Shortwave scores are for daylight profiles. Longwave scores pool all columns of each data set, and longwave heating is the temperature heating rate on the model levels. RMSE is the root-mean-square error, and a tendency is the rate of change that a scheme adds to the model state. A dash means that no bias was recorded.}
\label{S-tab:nn-offline}
\begin{tabular}{@{}L{0.27\textwidth}L{0.37\textwidth}rrr@{}}
\toprule
Development data & Quantity & RMSE & Bias & $R^2$\\
\midrule
\multicolumn{5}{@{}p{0.95\textwidth}@{}}{\textbf{Shortwave network (replaces Dudhia).} Learned: the absorbed, scattered and transmitted fractions of sunlight in each layer, from one network shared by all layers ($14\to32\to32\to3$, SiLU, 1,635 parameters). Kept: the incoming sunlight and the conversion of absorbed energy to a heating rate. Enforced: fractions that sum to one, energy passed down layer by layer and exactly zero heating at night.}\\
\addlinespace[2pt]
Generated profiles & Potential-temperature tendency (K\,day$^{-1}$) & 0.1566 & -- & 0.99819\\
 & Surface downward flux (W\,m$^{-2}$) & 6.681 & -- & 0.99946\\
 & Absorbed column flux (W\,m$^{-2}$) & 1.828 & -- & 0.99953\\
Cameron Peak & Potential-temperature tendency (K\,day$^{-1}$) & 0.04088 & -- & 0.99460\\
 & Surface downward flux (W\,m$^{-2}$) & 2.905 & -- & 0.99992\\
 & Absorbed column flux (W\,m$^{-2}$) & 0.308 & -- & 0.99962\\
\midrule
\multicolumn{5}{@{}p{0.95\textwidth}@{}}{\textbf{Longwave network (replaces RRTMG longwave).} Learned: the radiation response of a whole column (639 inputs, three 512-unit SiLU hidden layers, 378 outputs, 1,046,906 parameters). Kept: the input preparation and fixed gas amounts of the physical scheme. Enforced: nonnegative upward and downward fluxes, with heating computed from the change in net flux across each layer. Only this network saw Marshall columns in training, so its Marshall rows are not an independent test.}\\
\addlinespace[2pt]
Cameron Peak (27,648 columns) & Surface downward flux (W\,m$^{-2}$) & 5.457 & 1.051 & 0.9219\\
 & Heating rate (K\,day$^{-1}$) & 1.019 & $-0.102$ & 0.9726\\
Generated cold clear-sky (256 columns) & Surface downward flux (W\,m$^{-2}$) & 1.642 & 0.227 & 0.9974\\
 & Heating rate (K\,day$^{-1}$) & 0.302 & $-0.022$ & 0.7388\\
Generated cloudy (1,792 columns) & Surface downward flux (W\,m$^{-2}$) & 8.442 & 1.463 & 0.9925\\
 & Heating rate (K\,day$^{-1}$) & 1.422 & $-0.014$ & 0.6019\\
Marshall (8,960 columns) & Surface downward flux (W\,m$^{-2}$) & 9.961 & $-2.608$ & 0.8374\\
 & Heating rate (K\,day$^{-1}$) & 1.694 & 0.048 & 0.7055\\
\midrule
\multicolumn{5}{@{}p{0.95\textwidth}@{}}{\textbf{Spread network (replaces the Rothermel spread-rate factors).} Learned: the logarithms of the dimensionless base, wind and slope factors ($5\to64\to64\to3$, tanh, with a parallel linear connection from input to output, 4,757 parameters). Kept: fuel and moisture preprocessing, scaling to a dimensional spread rate, projection onto the direction of spread, rate caps, the non-burnable mask and the separate chaparral formula. No untouched test set was evaluated.}\\
\addlinespace[2pt]
16 development families & Spread rate, complete calculation (m\,s$^{-1}$) & 0.001415 & -- & 0.9999985\\
\bottomrule
\end{tabular}
\end{table}

The longwave network used in the forecasts includes the additional training on Marshall columns. Its Marshall scores are therefore not an independent test, although the Marshall development columns were held out from training. We selected this network under three conditions: errors on each original training group stay within tolerance, Cameron Peak heating RMSE stays below 1.1\,K\,day$^{-1}$, and cold clear-sky surface-flux bias stays within 3\,W\,m$^{-2}$. The longwave-only and all-three forecasts use the same longwave network. For the spread network, we report only development scores, and we evaluated no separate, untouched test set. The spread scores are for the complete spread calculation, in which moisture scaling and the directional calculations remain physical.

\subsection{Five single 24-hour forecasts}
\label{S-sec:emulator-coupled-scores}
The five single forecasts in this section are the ones mapped in Fig.~\ref{fig:emulators}a--c, and we show them as examples. Each is one member of a wide ensemble. Its difference from the physical forecast can therefore differ in size and sign from the ensemble-mean differences reported in the main text, which are the result we rely on (Section~\ref{S-sec:nn-numerical-sensitivity}). All five forecasts run stably for 24 hours with fire--atmosphere feedback and firebrands active, and each is compared with the same physical forecast (Supplementary Table~\ref{S-tab:emulator-twenty-four-hour}). Footprint IoU uses the cells where the level-set function is negative (Section~\ref{S-sec:fire-grid-registration}), and the model fire fields are neither shifted nor interpolated.

\begin{table}[htbp]
\centering\footnotesize\setlength{\tabcolsep}{4.5pt}
\caption{\textbf{The five single 24-hour forecasts with networks, compared with the physical forecast.} Area change and footprint IoU (the intersection over union of the neural and physical burned footprints) describe the fire. The other columns are RMS differences on the 111-m nest at 24 hours. Wind uses the lowest mass level, $T_{\rm skin}$ is the surface (skin) temperature, $H$ the sensible heat flux, and SW$\downarrow$ and LW$\downarrow$ the downward shortwave and longwave fluxes at the surface, all three in W\,m$^{-2}$. The last row gives the range across the four perturbed physical forecasts, each compared with the unperturbed physical forecast. Their burned areas span 211.57--333.05\,km$^2$, against 304.00\,km$^2$. All values are for single forecasts, shown as examples. The ensemble means are in the main text and Supplementary Table~\ref{S-tab:nn-ensembles}.}
\label{S-tab:emulator-twenty-four-hour}
\begin{tabular}{@{}lrrrrrrr@{}}
\toprule
 & \multicolumn{2}{c}{Fire} & \multicolumn{5}{c}{Atmosphere on the nest (RMS difference)}\\
\cmidrule(lr){2-3}\cmidrule(l){4-8}
Networks & Area change & IoU & Wind (m\,s$^{-1}$) & $T_{\rm skin}$ (K) & $H$ & SW$\downarrow$ & LW$\downarrow$\\
\midrule
Shortwave & $-6.98\%$ & 0.777 & 1.84 & 1.32 & 22.24 & 62.83 & 18.87\\
Spread & $+8.40\%$ & 0.844 & 1.57 & 1.08 & 18.75 & 48.73 & 13.33\\
Shortwave and spread & $+1.77\%$ & 0.793 & 1.74 & 1.38 & 21.84 & 60.03 & 16.87\\
Longwave & $+1.63\%$ & 0.711 & 1.85 & 1.48 & 22.84 & 54.39 & 23.24\\
All three & $-21.53\%$ & 0.681 & 1.87 & 1.77 & 27.10 & 62.12 & 21.50\\
\midrule
Four perturbed physical & $-30.4\%$ to $+9.6\%$ & 0.647--0.778 & 1.73--1.81 & 1.12--1.18 & 17.60--20.20 & 44.21--53.15 & 12.77--15.77\\
\bottomrule
\end{tabular}
\end{table}

As a scale for these single forecasts, we ran four perturbed physical forecasts. Each changes the initial winds by one unit in the last place, runs for 24 hours and is compared with the unperturbed physical forecast. This four-member set is separate from the ten-member ensembles of Section~\ref{S-sec:nn-numerical-sensitivity}. At 24 hours, the four members span 211.57--333.05\,km$^2$, compared with 304.00\,km$^2$ for the unperturbed forecast, and their footprint IoUs with that forecast span 0.647--0.778. In burned area and in footprint overlap, each of the five neural forecasts differs from the physical forecast by less than the largest of the four perturbed forecasts does.

The atmospheric fields differ more. The atmospheric columns of Supplementary Table~\ref{S-tab:emulator-twenty-four-hour} give RMS differences over the complete stored fields of the nest at 24 hours, with the staggered winds averaged to the cell centers. The shortwave, longwave and all-three forecasts exceed the four-member range in every column. The shortwave--spread forecast exceeds it for surface temperature, sensible heat flux and both downward radiative fluxes, and its wind difference lies inside the range. The spread-only forecast does not exceed the range in any column. The radiation networks change the radiative heating and the surface fluxes directly, so larger atmospheric differences are expected for the four forecasts that include one.

\subsection{Ten-member ensembles and six-hour comparisons}
\label{S-sec:nn-numerical-sensitivity}

We ran ten forecasts to 24 hours for each of six configurations: the physical forecast and five in which networks replace the shortwave scheme, the spread scheme, both, the longwave scheme, or all three. Each member changes initial wind components on both grids by one unit in the last place, the smallest possible change to a floating-point number. Every member reaches 24 hours. Table~\ref{S-tab:nn-ensembles} lists the mean, smallest and largest burned area in each ensemble.

\begin{table}[htbp]
\centering\small
\caption{\textbf{Burned area at 24 hours in the ten-member ensembles.} Members of an ensemble differ only by last-bit changes to the initial winds. The last column is the ratio of each ensemble mean to the physical ensemble mean, minus one.}
\label{S-tab:nn-ensembles}
\begin{tabular}{@{}lrrrr@{}}
\toprule
Configuration & Mean & Minimum & Maximum & Difference of mean\\
 & (km$^2$) & (km$^2$) & (km$^2$) & from physics (\%)\\
\midrule
Physics & 264.3 & 187.5 & 371.5 & --\\
Shortwave network & 272.1 & 196.4 & 356.9 & $+2.9$\\
Spread network & 274.8 & 175.1 & 372.1 & $+4.0$\\
Shortwave and spread networks & 307.0 & 264.7 & 357.2 & $+16.2$\\
Longwave network & 333.3 & 212.4 & 416.1 & $+26.1$\\
All three networks & 288.8 & 191.2 & 385.0 & $+9.3$\\
\bottomrule
\end{tabular}
\end{table}

Most members fall inside the range of the physical ensemble, and a few spread, longwave and all-three members fall outside it. The networks are demonstrations and were not tuned to match the physical ensemble. Percent changes in the main text are ratios of ensemble means. They can differ in size and sign from the single-forecast examples in Section~\ref{S-sec:emulator-coupled-scores}. Figure~\ref{fig:emulators} shows the ten-member ranges. Thirty paired forecasts at six hours, ten each for shortwave, spread and shortwave--spread, have mean burned areas that differ from the physical ensemble mean by $-0.44\%$, $-0.28\%$ and $+15.9\%$, respectively. The separate ten-pair longwave and boundary-layer (YSU, Section~\ref{S-sec:ysu-review}) comparisons give mean-area differences of $+11.3\%$ and $-8.4\%$, respectively, relative to the same physical ensemble.

\begin{figure}[htbp]
\centering\includegraphics[width=\textwidth]{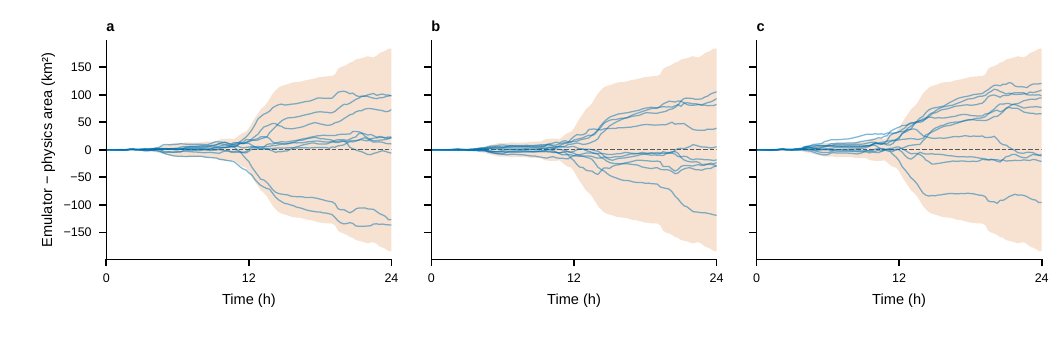}
\caption{\textbf{Paired ensemble responses.} Burned area of the neural forecast minus that of the physical forecast with the same initial perturbation, for ten pairs: shortwave (\textbf{a}), spread (\textbf{b}) and their combination (\textbf{c}). Curves use shared saved times without interpolation. The orange band spans all pairwise differences among the same ten physical members, $\pm(\max A_{\rm phys}-\min A_{\rm phys})$, and serves as a scale for numerical sensitivity. The dashed line marks zero.}
\label{S-fig:emulator-ensemble-paired}
\end{figure}

\clearpage

\section{Boundary-layer networks}
\label{S-sec:ysu-review}
An imperfect network can still complete a coupled forecast, and this section gives an example. The YSU scheme computes turbulent mixing in the boundary layer, the lowest part of the atmosphere. It runs only on the 1-km parent domain. We replaced the parts of YSU that estimate the boundary-layer height, the mixing coefficients, the non-local terms (mixing by large eddies that does not follow the local gradient) and the entrainment terms (mixing of air from above into the boundary layer) with two neural networks, each with three hidden layers of 1,024 units, 634 scaled inputs and bounded outputs. The numerical solution of the mixing equations in each column stays physical. The forecast never falls back to the physical formulas.

Training uses 414,660 columns (the first 80\% of column indices) from 20 saved times of the Marshall forecast. The spatial holdout, a set of columns kept out of training, is the last 10\% (51,840 columns) at those times. The intervening 10\% is not scored. The temporal holdout is all 233,253 columns at nine additional times. All groups come from one fire case. We scale the inputs with statistics from the training columns only. We first train the networks for 100 epochs (passes through the training data) to reproduce the quantities they replace, and then for five more epochs to reproduce the resulting tendencies (the rates of change that the scheme adds to the model state). Table~\ref{S-tab:ysu-fit} pools errors across columns, levels and saved times.

\begin{table}[htbp]\centering
\caption{\textbf{Accuracy of the YSU networks used in the forecast.} RMSE is in the stated physical units. $R^2=1-\mathrm{SSE}/\mathrm{SST}$, where SSE is the sum of squared errors and SST is the sum of squared deviations of the physical-scheme values from their pooled mean in each group. The first six rows are tendencies (rates of change that the scheme adds to the model state) and pool all 44 levels. Boundary-layer height is evaluated per column.}
\label{S-tab:ysu-fit}
\small\setlength{\tabcolsep}{4pt}
\begin{tabular}{lrrrrrr}\toprule
 & \multicolumn{2}{c}{Training} & \multicolumn{2}{c}{Spatial holdout} & \multicolumn{2}{c}{Temporal holdout}\\
Quantity (RMSE units) & RMSE & $R^2$ & RMSE & $R^2$ & RMSE & $R^2$\\\midrule
Temperature (K\,day$^{-1}$) & 13.62 & 0.921 & 36.41 & 0.675 & 24.39 & 0.629\\
Eastward wind (m\,s$^{-1}$\,day$^{-1}$) & 72.00 & 0.977 & 141.09 & 0.923 & 249.31 & 0.733\\
Northward wind (m\,s$^{-1}$\,day$^{-1}$) & 31.86 & 0.962 & 66.67 & 0.922 & 81.22 & 0.803\\
Water vapor (g\,kg$^{-1}$\,day$^{-1}$) & 5.24 & 0.774 & 10.40 & 0.387 & 10.30 & 0.264\\
Cloud water (g\,kg$^{-1}$\,day$^{-1}$) & 0.458 & 0.809 & 0.317 & 0.686 & 0.712 & 0.536\\
Cloud ice (g\,kg$^{-1}$\,day$^{-1}$) & 0.133 & 0.942 & 0.302 & 0.783 & 0.283 & 0.738\\
Boundary-layer height (m) & 55.22 & 0.991 & 93.38 & 0.975 & 196.15 & 0.862\\
\bottomrule\end{tabular}
\end{table}

The forecast completes all 28,800 parent steps with the networks called at every step and finite fields throughout. In tests of single calls, heat, water vapor and momentum are conserved in each column to within $3.05\epsilon S$, where $\epsilon$ is machine precision and $S$ sums the absolute column storage-rate and boundary-flux magnitudes. Conserving column budgets and running stably therefore do not guarantee accurate tendencies. Water-vapor tendency has the weakest fit at the withheld times ($R^2=0.264$), and temperature and eastward wind are also imperfect ($0.629$ and $0.733$).

\begin{figure}[p]\centering
\includegraphics[width=.98\textwidth]{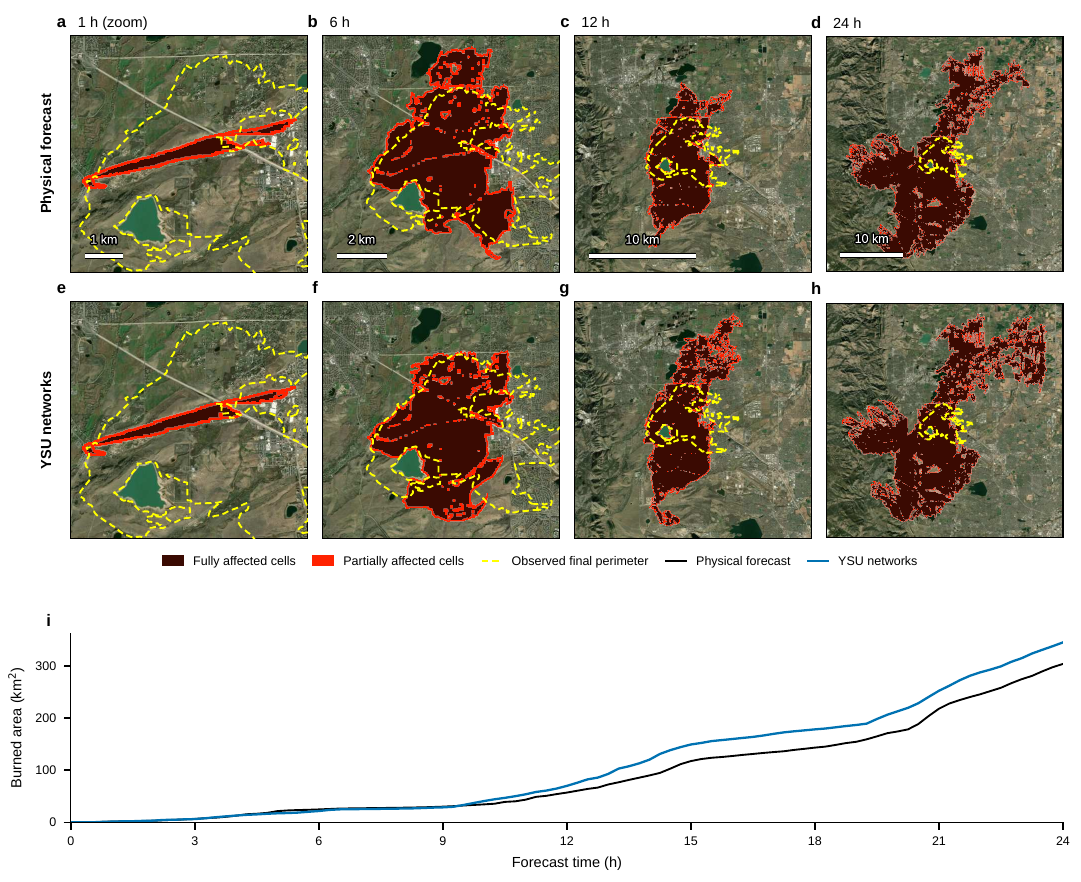}
\caption{\textbf{The YSU networks complete a 24-hour Marshall forecast.} \textbf{a--h}, Fully affected (dark red, burned fraction at least 0.999) and partially affected (bright red) fire cells in the physical forecast (a--d) and the forecast with the YSU networks (e--h) at one hour (zoomed), six, twelve and twenty-four hours. Yellow dashes mark the observed final perimeter. \textbf{i}, Burned area every fifteen minutes for the physical forecast (black) and the forecast with the YSU networks (blue). Nominal parent-step times are shown. Basemap, Esri World Imagery.}
\label{S-fig:ysu-coupled}
\end{figure}

The 24-hour fire area is 345.37\,km$^2$, compared with 304.00\,km$^2$ in the physical forecast, with a footprint IoU (intersection over union of the two burned footprints) of 0.759 (Fig.~\ref{S-fig:ysu-coupled}). Grid coordinates, fuel categories and saved times agree, but the neural and physical forecasts use different code versions, so this comparison does not isolate the effect of the networks. At 24 hours, all five atmospheric differences exceed the range of the four perturbed physical forecasts (Table~\ref{S-tab:ysu-numerical-context}). A separate ten-pair comparison with last-bit changes to the initial winds runs the same networks for six hours, and its mean burned area is 8.4\% below the physical ensemble mean (Section~\ref{S-sec:nn-numerical-sensitivity}).

\begin{table}[htbp]
\centering
\caption{\textbf{Atmospheric differences of the forecast with the YSU networks at 24 hours.} The YSU columns give the RMSE of the forecast with the YSU networks against the physical forecast, on the parent and on the nest. The two forecasts use different code versions, so these differences do not isolate the effect of the networks. Physical ranges are the smallest and largest values among the four perturbed physical forecasts, each compared with the unperturbed physical forecast. The two physical forecasts used as the basis of comparison are bitwise identical in the compared fields.}
\label{S-tab:ysu-numerical-context}
\begin{tabular}{lrrrr}\toprule
 & \multicolumn{2}{c}{Parent} & \multicolumn{2}{c}{Nest}\\
Quantity & YSU & Physical range & YSU & Physical range\\\midrule
Wind vector (m\,s$^{-1}$) & 1.58 & 0.31--0.36 & 2.07 & 1.73--1.81\\
Skin temperature (K) & 0.79 & 0.43--0.54 & 1.60 & 1.12--1.18\\
Sensible heat (W\,m$^{-2}$) & 12.93 & 8.22--10.95 & 29.01 & 17.60--20.20\\
Downward shortwave (W\,m$^{-2}$) & 36.16 & 22.49--29.58 & 79.47 & 44.21--53.15\\
Downward longwave (W\,m$^{-2}$) & 11.70 & 6.28--8.54 & 23.57 & 12.77--15.77\\
\bottomrule
\end{tabular}
\end{table}

\clearpage
\renewcommand{\refname}{References}
\bibliographystyle{naturemag}
\bibliography{references}

\end{document}